\documentclass[prd,showpacs,preprintnumbers,amsmath,amssymb]{revtex4}

\usepackage{graphicx}
\usepackage{dcolumn}
\usepackage{bm}

\newcommand{\NP}[1]{Nucl.\ Phys.\ {\bf #1}}

\newcommand{\PL}[1]{Phys.\ Lett.\ {\bf #1}}

\newcommand{\PR}[1]{Phys.\ Rev.\ {\bf #1}}
\newcommand{\PRL}[1]{Phys.\ Rev.\ Lett.\ {\bf #1}}

\newcommand{\Od}{{\cal O}}

\newcommand{\pint}{\int \frac{d^3 \vec{p}}{(2\pi)^3}}
\newcommand{\point}{\int \frac{d^3 \vec{p}_1}{(2\pi)^3}}
\newcommand{\ptint}{\int \frac{d^3 \vec{p}_2}{(2\pi)^3}}
\newcommand{\kint}{\int \frac{d^3 \vec{k}}{(2\pi)^3}}

\newcommand{\im}{\mbox{Im}\,}
\newcommand{\re}{\mbox{Re}\,}
\newcommand{\sgn}{\mbox{sgn}}

\newcommand{\fpi}{f_\pi}

\newcommand{\modq}{\vert \vec{q} \vert}
\newcommand{\modQ}{\vert \vec{Q} \vert}
\newcommand{\modp}{\vert \vec{p} \vert}
\newcommand{\modpo}{\vert \vec{p}_1 \vert}
\newcommand{\modpt}{\vert \vec{p}_2 \vert}
\newcommand{\modpsum}{\vert \vec{p}_1+\vec{p}_2 \vert}
\newcommand{\modpdif}{\vert \vec{p}_1-\vec{p}_2 \vert}
\newcommand{\modpq}{\vert \vec{p}+\vec{q} \vert}
\newcommand{\modpcapq}{\vert \vec{p}+\vec{Q} \vert}
\newcommand{\vq}{\vec{q}}
\newcommand{\vx}{\vec{x}}
\newcommand{\tsum}{T \sum_{n=-\infty}^{\infty}}

\newcommand{\be}{\begin{equation}}
\newcommand{\ee}{\end{equation}}
\newcommand{\ba}{\begin{eqnarray}}
\newcommand{\ea}{\end{eqnarray}}

\newcommand{\IZ}{{\Bbb Z}}
\newcommand{\IR}{{\Bbb R}}

\newcommand{\gsim}{\raise.3ex\hbox{$>$\kern-.75em\lower1ex\hbox{$\sim$}}}
\newcommand{\lsim}{\raise.3ex\hbox{$<$\kern-.75em\lower1ex\hbox{$\sim$}}}

\begin{document}

\title{The electrical conductivity of a pion gas}

\author{D. Fern\'andez-Fraile}
\email{danfer@fis.ucm.es}
\author{A. G\'omez Nicola}
\email{gomez@fis.ucm.es}

\affiliation{Departamentos de F\'{\i}sica Te\'orica I y II. Univ.
Complutense. 28040 Madrid. SPAIN.}

\begin{abstract}

The electrical conductivity  of a pion gas at low temperatures is
studied in the framework of Linear Response and  Chiral Perturbation
Theory.
 The standard ChPT power counting has to be modified to include pion propagator
 lines with a nonzero thermal width in order  to properly account for
  collision effects typical of Kinetic Theory. With this
  modification, we discuss the relevant chiral power counting  to
  be used in the calculation of transport coefficients.
 The leading order contribution is found and we show that the dominant
 higher order ladder diagrams can be treated as perturbative
   corrections at low temperatures. We find that the DC conductivity
  $\sigma (T)$ is a decreasing function of $T$, behaving for very low $T$  as
  $\sigma (T)\sim e^2m_\pi\sqrt{m_\pi/T}$, consistently with nonrelativistic Kinetic Theory.
   When unitarization effects are included, $\sigma(T)$
  increases  slowly as $T$ approaches  the chiral phase transition.
   We compare with related works and discuss  some physical consequences,
    especially in the context of     the low-energy hadronic photon
  spectrum in Relativistic Heavy Ion Collisions.

\end{abstract}


\maketitle

\section{\label{sec:intro} Introduction}

Transport coefficients provide relevant information about the
nonequilibrium response of a system to an external perturbation. In
particular, the electrical conductivity measures the response to
electromagnetic fields, the DC conductivity corresponding to
constant fields, i.e., to the limit of vanishing external frequency
and momentum. In this work we will be interested in the electrical
conductivity of a pion gas at low temperatures, as that formed after
a Relativistic Heavy Ion Collision has hadronized and cooled down
below the chiral phase transition. In this regime, the medium
constituents are predominantly light mesons, whose very low-energy
dynamics is described by Chiral Perturbation Theory (ChPT)
\cite{we79,gale84}. We remark that ChPT is a low-energy expansion
performed in terms of $p^2/\Lambda_\chi^2$ with $p$ a typical meson
energy or temperature and $\Lambda_\chi \sim 4\pi f_\pi\simeq$ 1.2
GeV and it has been successfully applied to light meson dynamics and
also to describe several aspects of the low-$T$ pion gas
\cite{gale87,gole89,schenk93,glp02,dglp02}.

In principle, ChPT should provide a meaningful expansion for any
thermal observable. However, as we will show below, transport
coefficients  like the electrical
 conductivity are intrinsically nonperturbative and require a
 suitable modification of the naive chiral power counting scheme to account for
  collision effects in the plasma, as dictated by Kinetic Theory.
 Similar features have been found  in the analysis of
 transport coefficients at high temperatures in weakly coupled theories \cite{jeon95,amy00,valle02}.
 In fact, from elementary Kinetic Theory \cite{landau} we expect the electrical
 conductivity to behave like $\sigma\sim e^2 N_{ch}\tau/m$  for a gas of
 charged particles of mass $m$ with density $N_{ch}$ and mean free time
 $\tau\sim 1/\Gamma$ with $\Gamma$ the particle width. Since in ChPT
 the pion width is formally $\Gamma=\Od(p^6)$
 \cite{gole89,schenk93} this indicates the need for considering
 non-leading corrections. In addition, the presence of
 nonperturbative insertions, typically of $\Od(1/\Gamma)$, could spoil the naive power counting of
 perturbation theory, requiring the sum of an infinite set of diagrams
 to get the correct proportionality constant for transport coefficients.
   This is indeed the case for high $T$ in
 scalar \cite{jeon95} and gauge \cite{valle02} theories, where it
 has been shown that the sum of certain ladder diagrams provide the
 dominant contribution and coincide with the leading order obtained
 using directly a Kinetic Theory approach \cite{amy00}. In Lattice
 Gauge Theory, there are  also estimates of the electrical
 conductivity for lower temperatures but still above the QCD phase
  transition \cite{gupta04}. As for other transport
 coefficients for the meson gas, the shear viscosity has been calculated in
  \cite{pra93,davesne96,dosa02,dolla04} in the Kinetic Theory framework.

 One of the main motivations of our work is then to analyze to what
  extent the previous observations apply at low temperatures in a
  pion gas, within the ChPT formalism. The fact that we are not dealing with a
  coupling constant expansion but with a low temperature and energy one will introduce
   important differences in the analysis of ladder diagrams.
    As a further interesting
  point, the DC electrical conductivity is directly related to the small
energy limit of the emission rate of photons from the plasma formed
after a Relativistic Heavy Ion Collision \cite{alam01}. The photon
emission rate is a very relevant observable, since the photons leave
the plasma almost without interaction, so that they carry important
information about  the Quark-Gluon-Plasma
\cite{aurenche98,amy01,gelis04,blage05} and
 the hadronic phase \cite{alam01,steele96,steele97,rawa99,turaga04}. Our analysis is in
principle applicable to get an estimate of the very low energy
photon rate steaming from the pion gas. In this context, we will
compare with related results in the literature and  with
experimental information.

The plan of the paper is the following: in section \ref{sec:form}
we will sketch the basic formalism used throughout this work, in
section  \ref{sec:dclo} the leading order for the DC conductivity
for low temperatures
 will be obtained, emphasizing the role played by the thermal
 width and the low-$T$ diagram power counting. Higher order
 diagrams which  could  be relevant  will be analyzed
 in section \ref{sec:higherorder}, where we will show in detail that for low
  temperatures one can treat those diagrams as perturbative
  corrections in the ChPT framework. The role of unitarity will be
  explored in section \ref{sec:unit} where we will also comment on
  possible consequences of our results in the context of the photon
  emissivity of the hadronic gas. Our conclusions are summarized in
  section \ref{sec:conc}. We have also included an Appendix with
  some  useful results for thermal loop correlators needed in the
  main text.

\section{Basic Formalism}
\label{sec:form}

We consider a pion gas at temperature $T$ fully thermalized, i.e,
with vanishing pion chemical potentials. The  response of this
system
 to an external classical electromagnetic field $A^\mu_{cl}$ can be
 described within the formalism of Linear
Response Theory, which gives the total thermal averaged
electromagnetic current to first order in the external field (Kubo's
formula) \cite{lebellac} as:

\be  <J^{tot}_{\mu} (\vec{x},t)> =-i \int d^4 x' A^{\nu}_{cl} (x')
\Delta^R_{\mu\nu} (x-x') \label{kubo1}
 \ee
 where
 $\Delta^R_{\mu\nu}(\vec{x},t)=\theta(t)<[J^{EM}_\mu(x),J^{EM}_\nu(0)]>$
 is the retarded current-current correlator.

 Here we will always treat photons as an external source, i.e, not thermalized
 with the pion gas. This is consistent with the idea that photons
 (and leptons) escape the plasma almost without interacting, since
 their interactions are weak. This assumption is commonly used when
 describing for instance dilepton or photon production, where the
electromagnetic part of the rate is always decoupled from the strong
contribution \cite{alam01}.  Therefore, and consistently with the
above, we will retain only the leading order contributions in the
electric charge. In other words, we are ignoring rescattering of
photons throughout the plasma.

With the  gauge choice of the external field $A^0_{cl}=0$, the
 spatial current in Fourier space is proportional to the classical electric
field:

\be <J^{tot}_{k} (\omega,\vq)>=\frac{\Delta^R_{k j}(\omega,
\vq)}{\omega} E_{cl}^j (\omega,\vq)\equiv \sigma_{kj}(\omega, \vq)
E_{cl}^j (\omega,\vq)  \ee where $\sigma_{kj}$ defines the
conductivity tensor. In position space:

\ba <J^{tot}_{k} (\vec{x},t)>&=&\int d^4 x'
K_{kj}(\vec{x}-\vec{x'},t-t') E^j_{cl} (\vec{x'},t')\nonumber\\
K_{ij}(\vec{x},t)&=&\int\frac{d\omega d^3\vq}{(2\pi)^4} \
\frac{\Delta^R_{ij}(\omega, \vq)}{\omega} \ e^{-i\omega
t}e^{i\vq\vec{x}}\label{condpos}\ea

We will be  interested here in the  DC conductivity, defined as the
linear response to a {\it constant} electric field, i.e.,
independent of time and space. In that case, one  defines the real
scalar conductivity $\sigma$ as $\re\sigma_{kj}=g_{kj}\sigma$ where:

\be \sigma=\frac{1}{6}\lim_{\omega\rightarrow
0^+}\lim_{\modq\rightarrow 0^+} \frac{\rho_k^k(\omega,
\modq)}{\omega}
\label{dcdef}\ee where $\rho=2\im (i\Delta^R)$ is the
current-current spectral function (see Appendix). In (\ref{dcdef})
it has been also used that, by Euclidean covariance,
  $(\Delta^R)_{ij}(\omega,\vec{0})=\frac{g_{ij}}{3}(\Delta^R)_k^k(\omega,\vec{0})$.
  We remark that the order in which the
 zero energy and momentum limits are taken in (\ref{dcdef}) are meant to ensure a physically
  meaningful answer and correspond to an electric field slightly nonstatic but constant in space.
  For instance, in the reverse order static charges would rearrange to give a
   vanishing electric current \cite{mahan}. However, the small energy and momentum limits
   are
   sometimes ill-defined as  showed below and then one has to be careful
   with their physical interpretation.

The DC conductivity is given  in terms of a retarded correlator.
We will work in the Imaginary Time  Formalism (ITF) of Thermal
Field Theory at finite temperature $T$ and calculate the
corresponding time-ordered products, which give the retarded Green
functions when the external frequencies $i\omega_n=2\pi n T$
($n\in\IZ$)  are continued analytically to continuous  values as
$i\omega_n\rightarrow \omega + i\epsilon$
\cite{lebellac,evansbaier}. We remark that the ITF has the
advantage of dealing with the same fields, vertices and diagrams
as the $T=0$ theory, with properly modified Feynman rules.
However, one has to make sure that the IT expressions obtained
after summing over internal Matsubara frequencies are analytic in
the external frequencies off the real axis. That will be the case
for all the quantities considered here, as showed  below. Our ITF
notation and methods for the calculation of transport coefficients
will follow closely those in \cite{valle02}. An alternative
procedure is to work within the Real Time Formalism (RTF) for
which the time contours include the real axis \cite{rt}.  On the
one hand, the number of degrees of freedom in the RTF doubles,
corresponding to the values of the fields above and below the real
axis and this complicates the diagrammatic evaluation. On the
other hand, the results in the RTF are already obtained for real
energies, so that no analytic continuation is needed and  retarded
Green functions can  be obtained from the time-ordered products by
simple diagrammatic rules \cite{kobes}. An example of a transport
coefficient  evaluated in the RTF can be found in \cite{wahe99}
for  the shear viscosity in the $\lambda\phi^4$ theory.

\begin{figure}[h]
\includegraphics[scale=.9]{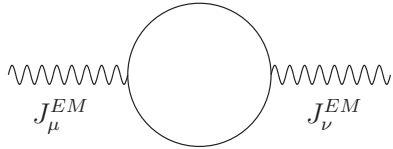}
 \caption{\rm \label{fig:diagLO} Leading order diagram
for the calculation of the DC electrical conductivity in ChPT
without collision effects.}
\end{figure}

\section{The DC conductivity in ChPT to leading order}
\label{sec:dclo}

\subsection{The ideal gas contribution}
\label{sec:dcideal}

Following the strict ChPT power counting, we should then start by
 considering the lowest order lagrangian $\Od(p^2)$ (the non-linear sigma
 model) coupled to an external photon. The lagrangian is given in
 \cite{gale84}. In the imaginary-time formalism we just take the
 same interaction vertices and electromagnetic current with the usual replacements for the
 Feynman rules, i.e, replacing all zeroth momentum components by discrete
frequencies $k^0\rightarrow i\omega_n=2\pi i n T$ and the loop
integrals  by Matsubara sums, i.e., $ \int \! \frac{dk^0}{2\pi}
\rightarrow i \tsum$.

The dominant diagram to the current-current spectral function is
that in Figure \ref{fig:diagLO}, which gives the following
imaginary-time contribution to the spatial current-current
correlation function :

\begin{widetext}
\be \left[\Delta_\beta\right]^0_{jk}(i\nu_m,\modq)=e^2\tsum\pint
\frac{(2p+q)_j (2p+q)_k}{\left[\omega_n^2+\modp^2+m_\pi^2\right]
\left[(\omega_n+\nu_m)^2+\modpq^2+m_\pi^2\right]}\label{currcorrlo}\ee
\end{widetext}
with $m_\pi\simeq$ 140 MeV the pion mass. For small external
momenta,  we can drop the $q$ dependence in the numerator, so that
we are led to the loop
 integral (\ref{loopcorr}) with $k=1$, analyzed in the Appendix.
  With the results showed there, one has  to
  this order:

\begin{widetext}
 \be \frac{\rho_k^k(\omega, \modq)}{\omega} (\omega\rightarrow
0^+,\modq\rightarrow 0^+)=\nonumber\\-\frac{e^2}{\pi}
\theta(\modq^2-\omega^2)
\frac{F_1\left[\omega^2/\modq^2;T\right]}{\modq}
\label{condnogamma}\ee
\end{widetext}
 with $F_1$ defined in (\ref{Fdef}). We
observe that the small energy and momentum  limits give an
ill-defined answer, due to the product of the vanishing denominator
and the step function (see our comments in the Appendix). In other
words, there are regions arbitrarily close
 to the origin in the $(\omega,\modq)$ plane where we get different
 limits for the spectral function: it vanishes for the timelike region
 $\modq^2<\omega^2 $ and grows with $1/\modq$ in the spacelike one $\modq^2>\omega^2
 $. Therefore, in this case instead of specifying a particular
 order in taking the limits,  we consider the
 Fourier transform of the contribution (\ref{condnogamma}) back to
 position space, according to (\ref{condpos}). We take
  electric fields slowly varying both in time and space, so that only
the small $\omega$ and $\modq$ limit of the spectral function is
relevant. When doing so, we get a divergent contribution for the
conductivity $\sigma=-\frac{e^2F_1(0;T)}{3\pi^2}\int_0^\infty dr$ to
leading order in the derivative expansion of the electric field,
where the dependence with the spatial length of the system is
explicitly shown. This is physically expected since, as commented in
the introduction, we are considering pions with zero width or
infinite mean free path. In fact, this  is nothing but the ideal gas
result (no collisions) since by considering only the diagram in
Figure \ref{fig:diagLO} we are not taking into account the effect of
pion interactions. Actually, to this order one would get the same
answer, say, in scalar QED. Therefore in an ideal gas, the charged
free pions travel throughout all the gas without interacting giving
an infinite conductivity. Thus, for an external field constant over
an infinite region, we should consider the effect of pion
interactions so that the DC conductivity is actually proportional to
the (finite) pion mean free path.  This will require a more careful
evaluation of ChPT diagrams, as we will show below.

\subsection{Beyond the ideal gas: finite pion width.}

\begin{figure*}\includegraphics[scale=.8]{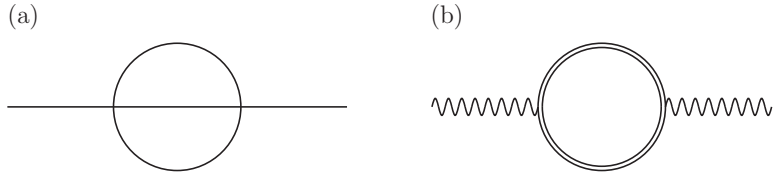}
 \caption{\rm
\label{fig:diagnlo}  (a) A two-loop diagram contributing to the pion
width. (b) Leading order current correlator with  dressed pion
lines.}
\end{figure*}

Our previous discussion makes it clear that in order to get finite
and physically sensible results for the DC conductivity, one has to
consider the effect of a finite pion width. In the medium, the pion
dispersion relation is $p^2=m_\pi^2+g(p_0,\modp;T)$ with $g$ a
complex function. The pion width is $\Gamma_p(\modp)=-\im
g(E_p,\modp)/(2E_p)<<E_p$ in ChPT with $E_p=\sqrt{\modp^2+m_\pi^2}$
and we will neglect the ChPT corrections to $\re g$, since they are
not relevant for our purposes.  The lowest order contribution to
$\im g$ in ChPT is $\Od(p^6)$ and can be calculated either by
Kinetic Theory arguments \cite{gole89} or by evaluating the relevant
diagrams \cite{schenk93} like the two-loop contribution shown in
Figure \ref{fig:diagnlo}a. The detailed results for $\Gamma_p$ can
be found in those references. For most of our purposes here it will
be enough to consider the dilute gas regime where $n_B(E_p)\approx
\exp(-E_p/T)<<1$:

\be \Gamma_p^{DG}=\frac{1}{2}\kint \sigma^{\pi\pi}_{tot} (s)
\frac{\sqrt{s(s-4m_\pi^2)}}{2E_k E_p}\exp (-E_k/T)\label{gammadg}\ee
where the Mandelstam variable
$s=(E_k+E_p)^2-\vert\vec{k}+\vec{p}\vert^2\geq 4m_\pi^2$ and
$\sigma^{\pi\pi}_{tot}$ is the total elastic pion-pion cross
section, that can be expressed as:

\be
\sigma^{\pi\pi}_{tot}(s)=\frac{32\pi}{3s}\sum_{J=0}^{\infty}\sum_{I=0}^2
(2J+1)(2I+1) \vert t_{IJ} \vert^2 \label{sigmatot}\ee in terms of
partial waves of definite isospin $I$ and angular momentum $J$, for
which we follow the same conventions for pion scattering as e.g. in
\cite{gale84}. To lowest order in ChPT, i.e, $\vert
t\vert^2=\Od(p^4)$:

\be \sigma^{\pi\pi}_{tot}(s)=\frac{5s}{48\pi f_\pi^4}
\left[1-\frac{16 m_\pi^2}{5s}+\frac{37
m_\pi^4}{10s^2}\right]\label{sigmatotlo}\ee where $f_\pi\simeq 93$
MeV is the pion decay constant. Since the integrand in
(\ref{gammadg}) is weighted by the Bose-Einstein factor, we expect
that taking the leading order for $\sigma^{\pi\pi}_{tot}$ is enough
for relatively low temperatures. We will comment below on the effect
of considering higher order terms, for instance those respecting
unitarity.

 The result (\ref{gammadg}) shows clearly how the pion traveling
 along the (dilute) medium with momentum $\vec{p}$ acquires a finite width by collisions
 with the medium constituents, the factor $\sqrt{s(s-4m_\pi^2)}/(2E_k
 E_p)$ being proportional to the relative velocity between the particles
 with momentum $p$ and $k$, expressed in relativistic form \cite{gole89}.

We consider then a pion propagator whose spectral function (see
Appendix) is:

\begin{widetext}
\ba \rho_\pi(\omega,\modp)=2  \im
iG^R(\omega,\modp)=\frac{\Gamma_p}{E_p}\left[\frac{1}{(\omega-E_p)^2+\Gamma_p^2}
-\frac{1}{(\omega+E_p)^2+\Gamma_p^2}\right] \ea
\end{widetext}

  Thus, the complex
propagator is:

\be G(z,\modp)=\frac{1}{E_p^2-\left[z+i\Gamma_p\sgn(\im z)\right]^2}
\label{complexG}\ee for $\im z\neq 0$, which has a cut on the real
axis and is analytic everywhere else in the complex $z$ plane.

We will denote by a double  line, the pion propagator ``dressed''
with a nonzero width $\Gamma_p$, so that the leading order
contribution to the current-current correlator is now given by
diagram (b) in Figure \ref{fig:diagnlo}. Its imaginary-time
contribution is now

\begin{widetext}
\be
\left[\Delta_\beta\right]_j^j(i\nu_m,\modq=0)=\nonumber\\4e^2\tsum\pint
\modp^2
G_\beta(i\omega_n,\modp)G_\beta(i\omega_n+i\nu_m,\modp)\label{cond0}\ee
\end{widetext}
where $G_\beta[i\omega_n]=G[i\omega_n]$ in (\ref{complexG}) for
$n\neq 0$ and $G_\beta(n=0)=1/(\Gamma_p^2+E_p^2)$ and where the
$\modq\rightarrow 0^+$ limit is now free of the ambiguities of the
ideal gas case. We will evaluate the Matsubara sum in (\ref{cond0})
following the same procedure as in \cite{valle02}. In this case, the
result can be expressed in terms of integrals along the two cuts of
$G$ at $\im z=0$ and $\im z=-\nu_m$ of the corresponding
discontinuities of the $G$ function. The result is an analytic
function for $i\nu_m$ off the real axis, so that we find for the
retarded correlator
($\Delta^R(\omega)=-i\Delta_\beta(i\nu_m\rightarrow
\omega+i\epsilon$)):

\begin{widetext}
\ba \left[\Delta^R\right]_j^j(\omega,\modq=0)&=&4e^2\pint \modp^2
\left\{G_\beta
(0)\left[G^R(\omega)-G^A(-\omega)\right]\right.\nonumber\\
&+&\int_{-\infty}^\infty \frac{dy}{2\pi} n_B(y)\left.
\left[G^R(y)+G^A(y)\right]
\left[G^R(y+\omega)-G^A(y-\omega)\right]\right\}\ea
\end{widetext}
where we have omitted the $\modp$ dependence in the propagators.
Now, it is not difficult to see that the leading order term in the
previous expression in the limits $\omega\rightarrow 0^+$ and
$\Gamma<<E_p$ is an $\Od(1/\Gamma)$ contribution coming from the
products $G^R G^A$ which in this limit behave like a $\delta$
function:

\be\int_{-\infty}^{\infty} dy G^R (y) G^A(y)
F(y)\simeq
\frac{\pi}{4 E_p^2
\Gamma_p}\left[F(E_p)+F(-E_p)\right]\label{pinchprod}\ee for
$\Gamma_p<<E_p$.

This type of ``pinching poles'' contribution is crucial for the
analysis of the conductivity. On the one hand, it is expected from
Kinetic Theory considerations, as discussed in the introduction. On
the other hand, it will lead to consider the effect of higher order
diagrams, as we will do in the following sections. Up to this point,
our result are analogous to those in \cite{jeon95,valle02} before
specifying the explicit form of $\Gamma_p$ and
  discussing the power counting. Similar ``pinching poles''
  contributions are obtained in real-time formalism calculations
  \cite{wahe99,amy01}.

  Thus, we get for the DC conductivity to leading order:

\begin{figure*}\includegraphics[scale=.6]{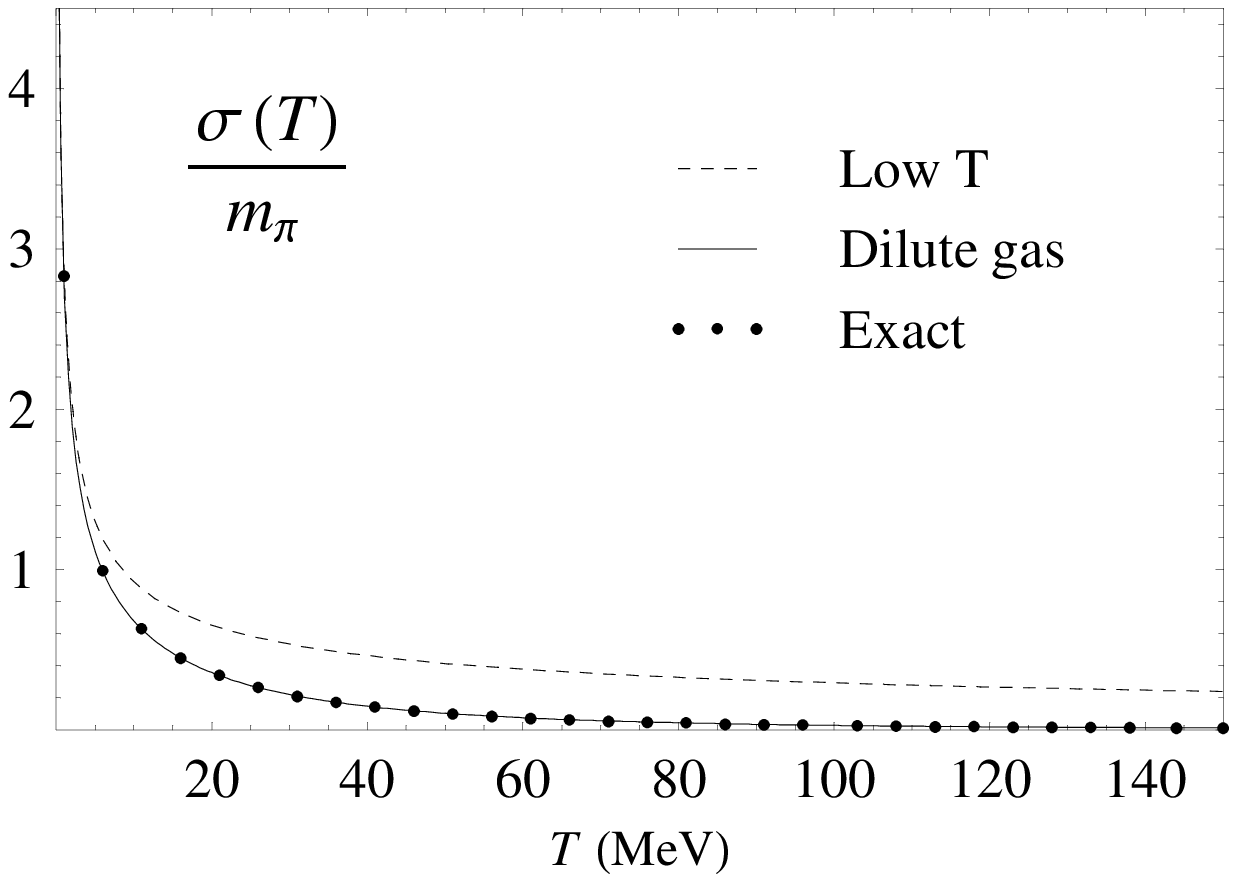}\includegraphics[scale=.6]{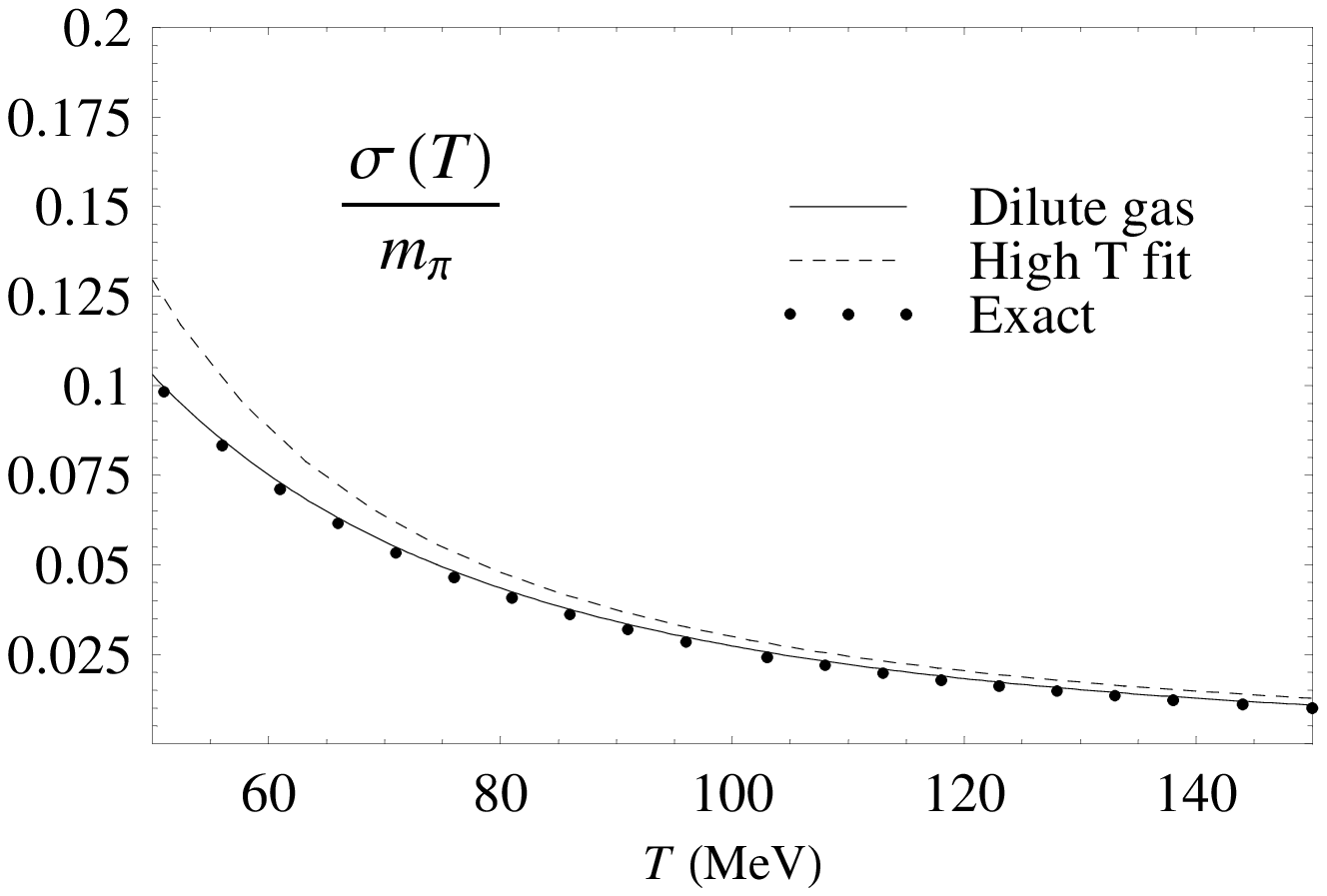}
 \caption{\rm
\label{fig:cond0}  Leading order for the DC conductivity. The left
panel shows the dilute gas approximation (solid line) compared to
the exact numerical result obtained  by equation (\ref{sigma0}) with
the full width calculated in \cite{gole89,schenk93} and to the
$T<<m_\pi$ limit (dashed line) given by (\ref{sigma0LT}). In the
right panel, the behaviour near $T\sim m_\pi$ is showed, where the
dashed
  line is given by equation (\ref{fitht}).}
\end{figure*}

\be \sigma^{(0)}=\frac{e^2}{3T}\pint \frac{\modp^2}{E_p^2 \Gamma_p}
n_B(E_p)\left[1+n_B(E_p)\right] \label{sigma0}\ee where we have used
that $n_B'(x)=-n_B(x)\left[1+n_B(x)\right]/T$ and
$n_B(-x)\left[1+n_B(-x)\right]=n_B(x)\left[1+n_B(x)\right]$. In
Figure \ref{fig:cond0} we have plotted the leading order
conductivity taking for $\Gamma_p$ the exact expression given in
\cite{gole89,schenk93} including all factors of $n_B$, comparing it
with the different approximations considered below.

The result (\ref{sigma0}) is very important  for our purposes. It
gives a finite and well-defined answer for the conductivity, with
the correct dependence on the pion width and the density as expected
from Kinetic Theory arguments. However, as commented before, the
nonperturbative nature of the DC conductivity, reflected in the
$1/\Gamma$ dependence, will make it necessary to examine higher
order diagrams.

Let us evaluate the previous result for the conductivity in the
dilute gas regime, which is valid for the range of temperatures of
interest here. Thus, we replace the results (\ref{gammadg}) and
(\ref{sigmatotlo})  in (\ref{sigma0}), where we neglect the $n_B^2$
term. The result is showed in Figure \ref{fig:cond0} (solid line)
where it is clear that the dilute gas regime in this case is an
excellent approximation for temperatures below the pion mass. The
conductivity of the pion gas grows for decreasing $T$, approaching
the ideal gas limit discussed in section \ref{sec:dcideal}. Note
that both the pion density  and the pion width vanish exponentially
for $T\rightarrow 0^+$, so that it was not clear what should be the
remaining $T$-dependence of their ratio. As we will show later, for
very low temperatures $T<<m_\pi$, $\sigma\sim m_\pi\sqrt{m_\pi/T}$.
This behaviour is nothing but the nonrelativistic limit predicted by
Kinetic Theory arguments and  indicates that there could be
nonperturbative contributions of higher order diagrams in the very
low $T$ region, where the typical $1/\Gamma$ insertions become
larger.

For this reason,  in the next section we will concentrate on the
$T<<m_\pi$ limit. As we shall see, in this limit the analysis
becomes simpler and our power counting arguments will be more
clearly illustrated.

\subsection{Very low temperatures}

In the $T<<m_\pi$ limit, the relevant integrals in the previous
expressions are dominated by the very low momentum region,
weighted by   $n_B(E_k)\approx \exp(-m_\pi/T) \exp (-k^2/2m_\pi
T)$. Thus, if we change $p\rightarrow \sqrt{m_\pi T} p$ in
(\ref{sigma0}) and $k\rightarrow \sqrt{m_\pi T} k$ in
(\ref{gammadg}), every power of a three-momentum integration
variable $\vec{p}/m_\pi$ in those integrals counts as $\sqrt{m_\pi
T}/m_\pi=\sqrt{T/m_\pi}$ and is therefore suppressed. Taking into
account that in this limit $\sqrt{s(s-4m_\pi^2)}\approx
2m_\pi\vert \vec{p}-\vec{k}\vert$ in (\ref{gammadg}), the angular
integration in the expression for the thermal width can be
performed and  the low-$T$ result can be expressed in terms of
error functions:

\be \Gamma_p^{LT}=\frac{ m_\pi^4 T^2}{E_p \fpi^4}\exp (-m_\pi/T)
  f(\hat p)\ee with

\be f(\hat p)= \frac{23}{512\pi^3} \left[\exp(-\hat
p^2/2)+\sqrt{\frac{\pi}{2}}\frac{1+\hat p^2}{\hat p} \mbox{Erf}
(\hat p/\sqrt{2})\right]\label{fdef}\ee and $\hat p=p/\sqrt{m_\pi
T}$. What is important from this expression is that for $E_p\approx
m_\pi$, $\Gamma_p\sim m_{\pi}^3 T^2/f_\pi^4\exp (-m_\pi/T)$ times  a
dimensionless function of $\hat p$ only so that we get readily the
dependence with $T,m_\pi,f_\pi$ of the thermal width for very low
temperatures,  as the product $\Gamma\sim nv\sigma^{\pi\pi}_{tot}$
with $n\sim (\sqrt{m_\pi T})^3$ (mean density) and
 $v\sim \sqrt{m_\pi T}/m_\pi$ (mean velocity) in the nonrelativistic
 limit of very small pion momentum compared to its mass. We recall
 that the same nonrelativistic behaviour is recovered at very low
 temperatures for the shear viscosity, which grows with $\sqrt{T}$
 \cite{dosa02,dolla04}.

Replacing the above result for the width in  the $T<<m_\pi$ limit of
(\ref{sigma0}) yields:

\begin{widetext}
\ba \sigma^{(0)}_{LT}= e^2 m_\pi\frac{f_\pi^4}{6\pi^2
m_\pi^4}\sqrt{\frac{m_\pi}{T}}\int_0^\infty d\hat p \frac{\hat
p^4}{f(\hat p)}\exp (-\hat p ^2/2)\simeq 2.7 e^2 m_\pi
\sqrt{\frac{m_\pi}{T}} \label{sigma0LT} \ea
\end{widetext}

Therefore, for low $T$ we get a $\sqrt{m_\pi/T}$ behaviour for the
conductivity in units of $m_\pi$.  This result
 will be crucial for our chiral power counting
analysis of higher terms in the DC conductivity and gives the
behaviour near the origin plotted in Figure \ref{fig:cond0} where we
compare (\ref{sigma0LT})  with the dilute gas approximation.

\subsection{Higher temperatures}

\label{sec:hit}

For temperatures up to the pion mass  we can trust the dilute gas
approximation, as shown in Figure \ref{fig:cond0}. On the other
hand, we do not expect ChPT to work much beyond that range. A fit to
the numerical results for temperatures around $m_\pi$ assuming a
single power form $aT^b$ gives:

\be \sigma^{(0)} (T)\simeq 0.82 \ e^2 m_\pi
\left(\frac{f_\pi}{m_\pi}\right)^4
\left(\frac{m_\pi}{T}\right)^{2.11} \qquad (T\lsim
m_\pi)\label{fitht}
 \ee

The above behaviour with $T$ can be understood as follows. Consider
the high temperature limit $T>>m_\pi$ in our previous expressions,
which  amounts to ignore the pion mass. Thus, for an integrand of
the form $f(p)  n_B(p)$, this is a better approximation for
functions $f$ dominated by their high energy behaviour. In fact, in
some cases  the high-$T$ curve is approached for not very high
temperatures, even when the dilute gas approximation still remains
valid and it is not a bad approximation to  neglect the pion mass in
the dilute gas expressions, in order to get, at least qualitatively
the behaviour for $T\lsim m_\pi$.  Consider for instance the total
density of charged pions $N_{ch}=(1/\pi^2)\int dk k^2 n_B(E_k)$. In
the $T>>m_\pi$ limit
 we replace $n_B(E_k)\simeq 1/(
\exp(k/T)-1)$ and obtain $N_{ch}\simeq 2\zeta(3) T^3/\pi^2=2.4
T^3/\pi^2$  whereas taking both the dilute gas and high-$T$
approximations, i.e. $n_B(E_k)\simeq \exp(-k/T)$, gives
$N_{ch}\simeq 2 T^3/\pi^2$, which around $T\sim m_\pi$ is only about
a 10$\%$ away from the exact curve. Something similar happens with
the pion width $\Gamma_p$, for which the high-$T$ limit in the
dilute gas expression (\ref{gammadg}) with the lowest order cross
section in (\ref{sigmatotlo}) gives $\Gamma_p\simeq 5 T^4 p/(12
\pi^3 f_\pi^4)$. Note that it is crucial for this argument that the
cross section grows with energy which in fact is not the physical
case, since unitarity bounds the partial waves. Therefore, we see
that implementing correctly the unitarity behaviour may change the
behaviour of the thermal width and hence of the conductivity for
intermediate and higher temperatures. We will examine this aspect in
section \ref{sec:unit}. Ignoring those possible corrections, the
mean thermal width $\bar \Gamma=\int \Gamma_p n_B(E_p)/\int
n_B(E_p)\simeq 5 T^5/(4\pi^3 f_\pi^4)$ in this limit, in agreement
with the analysis in \cite{gole89}. Therefore, we expect
qualitatively for the conductivity $\sigma\sim e^2
N_{ch}/(m_\pi\bar\Gamma)\sim e^2 m_\pi (f_\pi/m_\pi)^4 (m_\pi/T)^2$
as our above result (\ref{fitht}) roughly confirms. Performing  the
same approximations, i.e., dilute gas in the chiral limit, directly
in the expression (\ref{sigma0}) gives $\sigma^{(0)}=2\pi e^2
f_\pi^4/(5 T^3)$ which is worse than the fit (\ref{fitht}) at
$T\lsim m_\pi$, mostly because the integrand in (\ref{sigma0}) grows
much slower with $p$ than in the case of the width.

We also note here  that at higher temperatures, i.e, closer to the
chiral phase transition, one has to be careful with the approach we
are following throughout this work, since the thermal width becomes
comparable to the mass, as a consequence of the loss of validity of
ChPT. In fact, using our previous  high-$T$ expression for the width
gives $\bar \Gamma\simeq m_\pi/25$ at $T=$ 100 MeV but
$\bar\Gamma\simeq 1.2 m_\pi$ at $T=$200 MeV.

\section{Higher order diagrams}
\label{sec:higherorder}

Our previous analysis shows that the leading contributions to the DC
electrical conductivity come from nonperturbative $1/\Gamma$
insertions for every pair of pion lines carrying equal momentum as
the external four-momentum vanishes (double lines). Therefore, it is
not obvious that diagrams of higher order in the chiral power
counting are still suppressed with respect to the leading order
analyzed in the previous section.

We recall that according to Weinberg's power counting, the chiral
power of a given diagram with two external photon lines is $e^2 p^D$
with $D=2N_L+\sum_{d=2} (d-2)N_d$,  $N_L$ being the number of loops
and  $N_d$ the number of vertices coming from the lagrangian of
order $p^d$ in derivatives and meson masses. However, the presence
of ``pinching poles'' modifies this picture. Let us assign a
``nonperturbative'' factor $Y$ to the contribution of every two
internal double lines to the conductivity, so that we define
$\sigma^{(0)}=e^2 m_\pi Y$ and we call $X=\Od(p^2)$ a chiral power,
typically $X\simeq [m_\pi/(4\pi f_\pi)]^2$. Thus, the diagram in
Figure \ref{fig:diagnlo}b analyzed in the previous section is
$\Od(Y)$ instead of the $\Od(X)$ assigned by Weinberg's power
counting.

\begin{figure*}\includegraphics[scale=1]{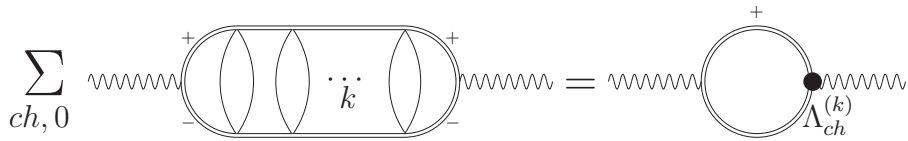}
 \caption{\rm
\label{fig:ladder} Ladder diagrams contributing to the electrical
conductivity.}
\end{figure*}

\begin{figure*}\includegraphics[scale=1]{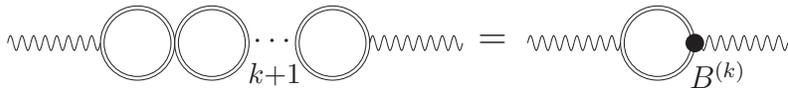}
 \caption{\rm
\label{fig:bubbles} A generic bubble diagram with $k+1$ bubbles
($k\geq 1$).}
\end{figure*}

The next task is then to identify those diagrams which are dominant
with this new power counting. In the case of a  scalar
$\lambda\phi^4$ theory, one has \cite{jeon95} $Y=1/\lambda^2$ and
the dominant contributions  are ladder diagrams of the type showed
in Figure \ref{fig:ladder}, which are all $\Od(1/\lambda^2)$ and
have to be resummed, giving the results expected from Kinetic
Theory. Other diagrams which naively could give  larger
contributions, like the ``bubble'' diagrams showed in Figure
\ref{fig:bubbles} have been shown to be subdominant. The same
conclusion about the dominance of ladder diagrams holds in gauge
theories \cite{amy00,valle02,amy01}.

The arguments based on the topology of the diagrams in the scalar
theory apply directly to our case, so that we are led to examine the
same class of   diagrams. However, it must be pointed out that in
ChPT there are important differences: first, we are actually dealing
with a double counting in $X$ and $Y$ as explained before, which is
a consequence of the importance of chiral loops in the power
counting. Second, we are allowed to have derivative vertices and
multiple vertices coming from lagrangians of different orders. And
finally, there are two types of particles, charged and neutral,
running in the internal lines and loops. This motivates our detailed
analysis below, which main conclusions will be that here the ladder
diagrams also dominate over other diagrams like the bubble ones, but
their influence  is not as important as it could be and can be
safely considered subdominant with respect to the leading order
result, within the ChPT validity range.

Let us start by considering  then the ladder diagrams showed in
Figure \ref{fig:ladder} where all the vertices come from the $d=2$
lagrangian. Note that only the internal lines running along the
ladder can carry equal momentum when the external momentum
vanishes, so that we keep the double line notation for those lines
and the single line for the rung loops. Our
 power counting gives then  $\Od(X^n Y^{n+1})$ for the diagram with
 $n$ rungs. Therefore, those diagrams could in principle be of the
 same or even higher order than the zeroth order diagram analyzed in
 the previous section, depending of the relative size between $X$ and $Y$.
  Now recall that according to our previous
 analysis, the nonperturbative $Y$ factors are bigger the smaller
 the temperature, where typically $Y\sim \sqrt{m_\pi/T}$. It is
 therefore at very low $T$ where we have to be  more concerned
 about these contributions. The detailed analysis in this regime is done in  the next
 section, where we will see that at low $T$ one has to account also for the Bose-Einstein
 distributions emerging from the loops, which will give rise to additional suppressions.

\subsection{Ladder diagrams at very low $T$}
\label{sec:laddlowt}

For the analysis of the generic ladder diagram showed in Figure
\ref{fig:ladder} we will follow similar steps as in \cite{valle02}.
The contribution of the diagram with $k$ rungs to the
current-current correlator is now:

\begin{widetext}
\ba
\left[\Delta^{(k)}_\beta\right]_j^j(i\nu_m,\vec{0})=e^2\tsum\point
2\vec{p}_1\cdot\vec{\Lambda}_{ch}^{(k)}(i\omega_n+i\nu_m,i\omega_n,\vec{p}_1)
\nonumber\\\times G_\beta(i\omega_n,\modpo)
G_\beta(i\omega_n+i\nu_m,\modpo) \label{condladdern}\ea
\end{widetext} where $\vec{\Lambda}_{ch}^{(k)}$ is the effective $\pi^+\pi^-\gamma$
vertex made of $k$ rungs, as depicted in Figure \ref{fig:ladder},
where the sum is made over all possible ways of inserting internal
lines and vertices corresponding to charged and neutral pions,
respecting  charge conservation and including the corresponding
combinatoric factors. Note that the two pairs of lines attached to
the external lines are always charged, denoted in the figure by
$+,-$ symbols. The lowest order effective vertex is
$\vec{\Lambda}_{ch}^{(0)}=2\vec{p}_1$.

The effective vertices satisfy a recurrence relation which
ultimately leads to integral equations corresponding to linearized
Boltzmann equations \cite{jeon95,valle02}. The same equations for
the effective vertices are obtained in the real-time formalism
\cite{wahe99,amy01}. In our present case, the integral equations
become a little more involved due to the presence of neutral and
charged particles, as well as derivative vertices. However, for
$T<<m_\pi$ an important simplification takes place: The leading
order of the sum in Figure \ref{fig:ladder} for a given number of
rungs $k$ is proportional to a single diagram where all vertices are
replaced by constant vertices $m_\pi^2/f_\pi^2$ and therefore the
spectral function of the rung integral is given by that of the loop
correlator $J_0$ in (\ref{loopcorrj}), analyzed in the Appendix.

Let us justify our previous statement. We have seen in the previous
section that the presence of a pair of double lines forces its
momentum to be on shell at $p_0=\pm E_p$. Thus, consider the
one-loop subgraph made of two ``incoming'' double lines and two
``outgoing'' ones. The two vertices contain different combinations
of powers of  two momenta or  two masses in each vertex, where the
momenta can be ``external'' (double lined) $p_1,p_2$ or ``internal''
$k$ and $k+p_1\pm p_2$ (see details below). In addition, as we will
see below, the integrals over $p_1$ and $p_2$ contain Bose-Einstein
factors that, according to our very low $T$ counting, suppress
powers of $\vec{p}_{1,2}/m_\pi$. Therefore, the leading contribution
for $T<<m_\pi$ in  $m_\pi^4$ or $p_{1,2}^4$  vertices is
proportional to $m_\pi^4 J_0(k)$, since $E_{1,2}\simeq m_\pi$. As
for the powers involving ``internal'' momenta,  using that
$E_1+E_2\simeq 2m_\pi>>\vert
\vec{p_1}-\vec{p_2}\vert>>E_1-E_2=\Od(\modp_{1,2}^2/m_\pi)$ and the
relationship among the different one-loop thermal integral with
momentum powers in the numerator discussed in Appendix A of
\cite{glp02}, we have checked explicitly that we are left, to
leading order, only with $m_\pi^4 J_0$, $m_\pi^3J_1$ or $m_\pi^2J_2$
contributions to the imaginary part (spectral function) of the
loops, which,  as our analysis below will show, provides the
relevant contributions to the ladder diagrams.  The imaginary part
of the $J_\alpha$ integrals is given in the Appendix. For $T<<m_\pi$
we can replace the Bose-Einstein distributions by Boltzmann
exponentials and the relevant integrals can be explicitly evaluated,
with the result that both for the thermal and for the unitarity
cuts, the leading order of  $\im J^R_\alpha$ are proportional  to
$m_\pi^\alpha \im J^R_0$ for $\alpha=1,2$.

\begin{figure}\includegraphics[scale=1]{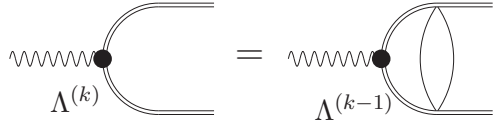}
 \caption{\rm
\label{fig:recurr} Recurrence relation between ladder vertices.}
\end{figure}

Therefore, for $T<<m_\pi$ we have

\be\vec{\Lambda}_{ch}^{(k)}=\left(\frac{m_\pi}{f_\pi}\right)^{4k}\alpha^{(k)}\vec{\Lambda}^{(k)}\ee
with $\alpha^{(k)}$ a purely numerical factor and
\begin{widetext}
\ba
\vec{\Lambda}^{(k)}(i\omega_m+i\nu_m,i\omega_m,\vec{p}_1)&=&\int_{-\infty}^{\infty}
\frac{d\omega'}{2\pi}\ptint \rho(\omega',\vert
\vec{p}_2-\vec{p}_1\vert)\tsum\vec{\Lambda}^{(k-1)}
(i\omega_n+i\nu_m,i\omega_n,\vec{p}_2)\nonumber\\
&\times& \frac{G_\beta(i\omega_n,\modpt)
G_\beta(i\omega_n+i\nu_m,\modpt)}{\omega'-i(\omega_n-\omega_m)}\label{evlt}\ea
\end{widetext}
 where
$\rho(\omega,\modq)=2\im J_0^R(\omega,\modq)=2\im
I^R(\omega,\modq;0)$ according to our notation in the Appendix. The
above recurrence relation is depicted in Figure \ref{fig:recurr} and
allows us to proceed along similar steps as in \cite{valle02}. By
induction, we have that the only singularities of
$\vec{\Lambda}^{(k)}(z+i\nu_m,z,\vec{p})$ are the same two cuts as
the propagator product in (\ref{evlt}), i.e, at $\im z=0$ and at
$\im z=-\nu_m$. This allows to perform the frequency sum in
(\ref{condladdern})
 in a similar manner as we did in the previous section and
  also the sum in the equation for the effective
vertex (\ref{evlt}), where in the latter case  we have to
consider, in addition to the cuts contribution, the pole
contribution at $z=\omega'+i\omega_m$. As in \cite{valle02}, the
previous arguments on the analytic structure of the effective
vertex imply that the analytic continuation $i\nu_m\rightarrow
\omega+i\epsilon$ is well defined. Altogether, we find for the
 low-$T$ conductivity
 to leading order in $1/\Gamma$, using (\ref{pinchprod}):

\begin{widetext}
\be \sigma^{(k)}=\frac{e^2
\alpha^{(k)}}{3T}\left(\frac{m_\pi}{f_\pi}\right)^{4k}\point
\frac{\modpo^2}{E_1^2 \Gamma_1} n_B(E_1)\left[1+n_B(E_1)\right]
A^{(k)}(\modpo)\label{sigmanlt}\ee with \ba A^{(k)}(\modpo)&=&
\frac{1}{\modpo^2}\ptint \frac{\vec{p}_1\cdot\vec{p}_2}{8 E_2^2
\Gamma_2}A^{(k-1)}(\modpt) \left\{
\left[n_B(E_2-E_1)-n_B(E_2)\right]\rho\left(E_2-E_1,\vert
\vec{p}_2-\vec{p}_1\vert\right)\right.\nonumber\\
 &-&\left.
\left[n_B(E_2+E_1)-n_B(E_2)\right]\rho\left(E_2+E_1,\vert
\vec{p}_2+\vec{p}_1\vert\right)\right\}\label{Anlt}\ea
\end{widetext}
where we have also used that $\rho(\omega')=-\rho(-\omega')$ and we
have denoted for the effective vertex
$\vec{\Lambda}^{(k)}(x+i\epsilon,x-i\epsilon,\vec{p})=2\vec{p}\tilde
A^{(k)}(\modp;x)$
 and $A^{(k)}(\modp)=(\tilde A^{(k)}(\modp;E_p)+\tilde A^{(k)}(\modp;-E_p))/2$ so that
 $A^{(0)}=1$. For simplicity, we have  denoted $E_{1,2}=E_{p_1,p_2}$ and so
on for $\Gamma_{1,2}$ and omitted their explicit  dependence with
$\modpo,\modpt$.

Let us comment on the result (\ref{Anlt}). The two terms in the
right hand side  correspond to evaluate the imaginary part of the
rung loop at $p_1\pm p_2$ for on-shell $p_{1,2}$ ``external''
momenta. That is, those contributions arise from the $s,t$ channels
of the
 $\pi\pi$ elastic scattering amplitude of the double lines
in Figure \ref{fig:recurr}, to one loop. Here we call by convenience
${\cal S}\equiv (E_1+E_2,\modpsum)$ and ${\cal T}\equiv
(E_1-E_2,\modpdif)$. This interpretation will be very useful in our
analysis of the different ChPT diagrams. In fact, note that the
remaining $u$-channel contribution would be given by the
contribution to the effective vertex of the
 bubble diagrams in Figure \ref{fig:bubbles}, since the
 loop integral of the  bubble does not depend on $p_1$ and
 $p_2$, only  on the external photon momentum $q$, i.e, we can identify
 ${\cal U}\equiv(\omega,\vec{0})$. Now  we remark (see Appendix) that the imaginary part
$\rho(E,\modQ)$ is nonzero both
 for $E^2\geq \modQ^2+4m_\pi^2$  (``unitarity'' cut) and
 for $E^2\leq \modQ^2$ (``thermal'' cut). Since $s={\cal S}^2\geq
 2m_\pi^2$,  the $s$-channel is given by the unitarity contribution
 in (\ref{unitcut0}) with $E=E_1+E_2,
 \vec{Q}=\vec{p}_1+\vec{p}_2$, while $t={\cal T}^2\leq 2m_\pi^2$ and
 thus the $t$-channel is given by the thermal part in
 (\ref{thercut0}) with $E=E_1-E_2,
 \vec{Q}=\vec{p}_1-\vec{p}_2$. Note that a consequence of this
 analysis is that, as announced, the bubble diagrams in Figure \ref{fig:bubbles}
 are indeed suppressed in the $\omega\rightarrow 0^+$ limit since
 $u={\cal U}^2=\omega^2$ does not fall within any of the two cuts.
 This  will be confirmed in section \ref{sec:other}.

At very low $T$, there are further simplifications of the above
equations, namely
$n^2(E_{1,2}),n(E_1+E_2)<<n(E_{1,2})\simeq\exp(-E_{1,2})/T$.
Consistently, in the ${\cal S}$-channel, we can neglect the
$T$-dependent part in the unitarity contribution (\ref{unitcut0}).
In the ${\cal T}$ channel, we can approximate in (\ref{thercut0})
the logarithm in the right hand side by $e^{-y_0/T}\left(1-e^{-\vert
E_1-E_2 \vert}\right)=e^{-(y_0+\vert E_1-E_2 \vert)/T}/n_B(\vert
E_1-E_2 \vert)$. Since $y_0\geq m_\pi$, the leading contribution in
that term comes from the $n_B(E_2-E_1)$ term in (\ref{Anlt}), giving
a net $\Od(e^{-2m_\pi/T})$ contribution, the same as in the ${\cal
S}$-channel and in the $\Gamma_1\Gamma_2$ in the denominator.
 The latter  approximations are valid in the dilute gas regime,
 which as commented previously, extends to $T\lsim m_\pi$. In
 addition, for $T<<m_\pi$  in the ${\cal S}$-channel, the
 product of the Bose-Einstein distributions $n(E_1)n(E_2)$
 implies $E_{1,2}\simeq
 m_\pi+[\vert \vec{p}_{1,2}\vert^2/(2m_\pi)]$ and hence $\sigma_{2p}\simeq
 \modpdif/(2m_\pi)$. Finally, we have  for
 $T<<m_\pi$:

\begin{widetext}
\begin{eqnarray} \sigma^{(k)}_{LT}=e^2\alpha^{(k)}\frac{m_\pi}{3}
\left(\frac{f_\pi}{m_\pi}\right)^{4} \sqrt{\frac{m_\pi}{T}}\int
\frac{d^3\vec{y}_1}{(2\pi)^3}\frac{y_1^2}{f(y_1)}
A_{LT}^{(k)}\left(\sqrt{m_\pi T} y_1\right)
\exp\left(-\frac{y_1^2}{2}\right)\label{sigmaklt}\\\hfill\nonumber\\
A_{LT}^{(k)}\left(\sqrt{m_\pi T} y_1\right)=\frac{1}{32\pi y_1^2}
\int
\frac{d^3\vec{y}_2}{(2\pi)^3}\frac{\vec{y}_1\cdot\vec{y}_2}{\vert
\vec{y}_1-\vec{y}_2\vert f(y_2)} A_{LT}^{(k-1)} (\sqrt{m_\pi T}
y_2)
\exp\left(-\frac{y_2^2}{2}\right)\nonumber\\
\nonumber\\\times\left\{\frac{\vert \vec{y}_1-\vec{y}_2\vert^2}{4} +
\exp\left[\frac{1}{4}\left(
y_1^2+y_2^2-\frac{1}{2}\frac{(y_1^2-y_2^2)^2}{\vert
\vec{y}_1-\vec{y}_2\vert^2}-\frac{1}{2}\vert
\vec{y}_1-\vec{y}_2\vert^2 \right)\right]\right\}\nonumber\\
\label{aklt}
 \end{eqnarray} \end{widetext}
 for $k\geq 1$, where the function $f$ is defined in (\ref{fdef})  and we have rescaled
$\vec{p}_{1,2}=\sqrt{m_\pi T} \vec{y}_{1,2}$ ($y_{1,2}\equiv\vert
\vec{y}_{1,2}\vert$).

From the previous equations (\ref{sigmaklt})-(\ref{aklt}) we can
draw one of our main conclusions: the low-$T$ effective vertex
$A_{LT}^{(k)}\left(\sqrt{m_\pi T}y_1\right)$ is $T$-independent
and hence, from (\ref{sigmaklt}) the correct order for the
contribution of the $k$-rung ladder to the conductivity in our
power counting scheme is $\sigma_{LT}^{(k)}=e^2 m_\pi \Od( X^k
\sqrt{m_\pi/T} )=e^2 m_\pi \Od(X^k Y)$ rather than the  $\Od(X^k
Y^{k+1})$ following from direct inspection, as discussed at the
beginning of this section. The independence of the LT effective
vertex on $T$ follows from (\ref{aklt}) by induction, since
$A^{0}=1$. Then, the only dependence with $T$ in
$\sigma_{LT}^{(k)}$ is the $1/\sqrt{T}$, the same as the lowest
order (\ref{sigma0LT}).  We can also understand why the ``direct''
counting misses $k$ inverse powers of $Y$. The reason is that the
contribution from the spectral functions  introduce an extra
$\sqrt{T/m_\pi}$ factor inside the integrand in (\ref{aklt}).
These factors come from $\rho_{\cal S}\sim \vert
\vec{p}_1-\vec{p}_2 \vert /m_\pi$ and $\rho_{\cal T}\sim T/\vert
\vec{p}_1-\vec{p}_2 \vert$ before rescaling the integration
variables. This  reflects the fact that the imaginary parts are
small at low $T$ due to the presence of the Bose-Einstein factors
and represents one of the main differences between our analysis
and the standard one in a coupling constant perturbation theory.
 The remaining $\Od(X^k)$ contribution
comes from having $k$ chiral loops. To check explicitly our
conclusions, we have evaluated numerically the one-rung
contribution $k=1$ in (\ref{sigmaklt})-(\ref{aklt}) setting
$\alpha^{(1)}=1$. We obtain
$\sigma_{LT}^{(1)}/\sigma_{LT}^{(0)}\simeq 0.06$. The reason why
this gives a slightly higher value than expected by power counting
is the presence of a collinear effect in the ${\cal T}$-channel,
coming from the region $\vec{y}_1\simeq \vec{y}_2$
  in (\ref{aklt}). The integral is convergent but that region
  enhances its value with respect to the ${\cal S}$ channel. In
  fact, for $k=1$ the ${\cal T}$-channel contribution is about five times
  larger than the ${\cal S}$-channel one, which has the expected chiral reduction mentioned
  above. Collinear effects of this kind  are characteristic of
   these analysis and  play  a crucial role in gauge
   theories \cite{moore04,aurenche98}. In our case, this numerical enhancement does not spoil the
   perturbative nature of these contributions.

   We can then conclude that the ladder diagrams at very low $T$, although
   representing the main contribution beyond leading order, are perturbative in
   the ChPT scheme.   In numerical terms, the very low $T$ conductivity is
   given by (\ref{sigma0LT}) with a theoretical uncertainty of
   only about  5\% in the multiplying coefficient.

\subsection{Ladder diagrams for $T\lsim m_\pi$}
\label{sec:laddhight}

Our previous analysis shows that the ladder diagrams are still
chirally perturbative in the very low $T$ region, even though we
should not miss the fact that their correct counting gives a much
larger contribution than their standard ChPT one, namely
$\sigma^{(k)}/\sigma^{(0)}=\Od(p^{2k})$ instead of the $\Od(p^{4k})$
given by Weinberg's theorem. The reason is the presence of the
nonperturbative $Y$ factors, as we have seen. Following our chiral
counting arguments for temperatures of the order of $m_\pi$, we
realize first that now $Y\sim \Od(1)$ around $T\sim m_\pi$. This
means that we expect again $\sigma^{(k)}/\sigma^{(0)}=\Od(p^{2k})$
for the $k$-ladder diagram, coming from the rung loops. It must be
reminded though that for these temperatures, the simplification of
reducing any ladder diagram to one with constant vertices does not
hold. In fact, for higher temperatures we expect derivative vertices
to become increasingly important, but as long as $T$ remains within
the ChPT applicability range (not much above the pion mass) we can
consider all those diagrams perturbatively suppressed.

\begin{figure*}\includegraphics[scale=.65]{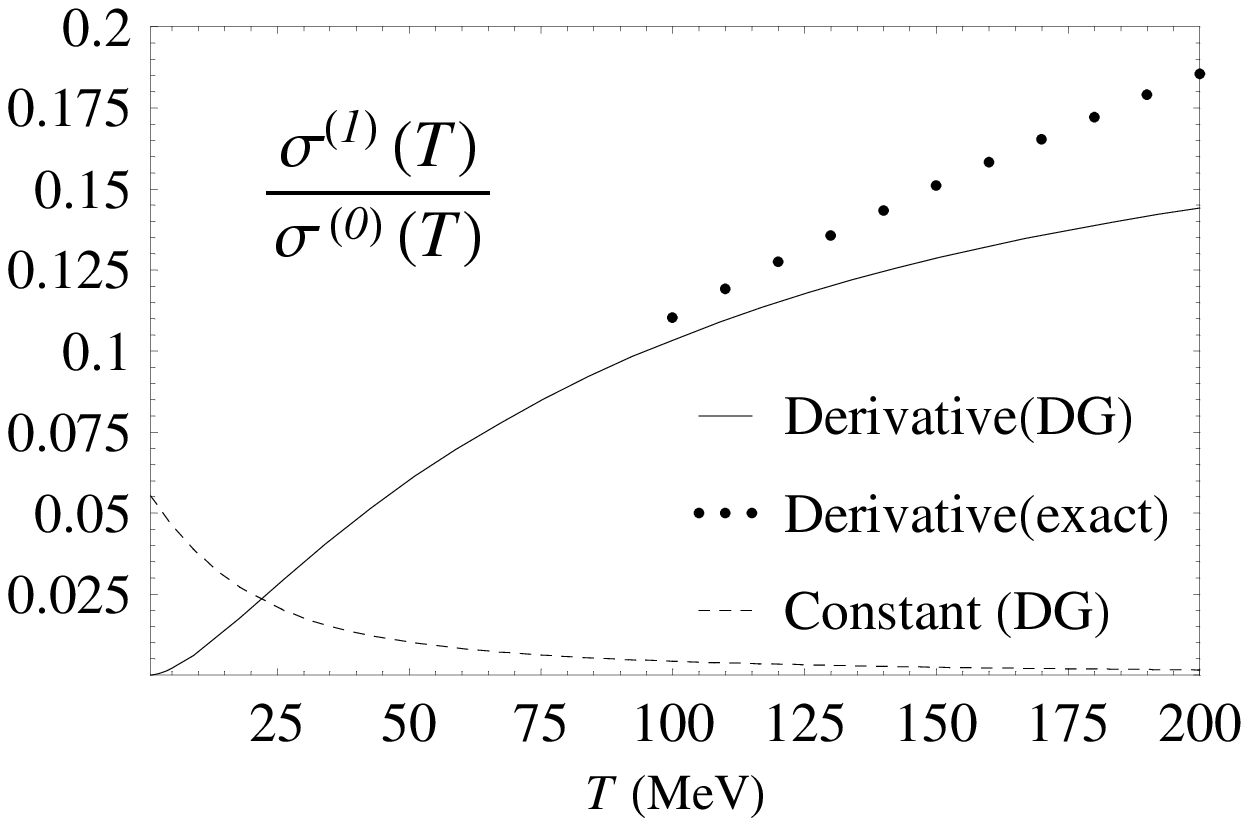}
 \caption{\rm
\label{fig:htcomp}  Estimation of the contribution of ladder
diagrams for higher temperatures, with and without derivative
vertices. The points correspond to the exact expressions without
using the dilute gas approximation.}
\end{figure*}

To check our previous comment we have evaluated numerically the
ladder diagram with one rung and constant vertices, setting in
(\ref{sigmanlt})-(\ref{Anlt}) $k=1$ and $\alpha^{(1)}=1$ and
performing only the dilute gas approximations mentioned in the
previous section but not those specific of $T<<m_\pi$. We have also
considered, for comparison, the contribution of a ladder diagram
with one rung but with one possible combination of derivative
vertices, namely, that obtained by taking $k=1$ in
(\ref{sigmanlt})-(\ref{Anlt}) but replacing the constant
$(m_\pi/f_\pi)^4$ factor by $\vert \vec{p}_1\vert^4/f_\pi^4$ in the
integrand. The results are showed in Figure \ref{fig:htcomp}, where
we have showed for comparison some points calculated without using
the dilute gas approximation. Clearly, the derivative vertices
become the more important ones at moderate and high temperatures,
showing that the ladder diagrams have to be summed at temperatures
close to the phase transition, where the ChPT power counting fails.

\subsection{Other diagrams}
\label{sec:other}

Following our previous arguments, it is not difficult to conclude
that, as far as the ladder diagrams analyzed in sections
 \ref{sec:laddlowt} and \ref{sec:laddhight}
 are perturbatively
 controlled in the chiral expansion, so are the same diagrams  with
 vertices coming from the lagrangians ${\cal L}_d$ with $d>2$ and also ladder diagrams
  with rung loops made with more than four pion vertices.

As for the bubble diagram showed in Figure \ref{fig:bubbles} which
could in principle give nonperturbative contributions of
$\Od(Y^n)$, our previous discussion based on the scattering of
internal lines has already suggested that these diagrams do not
receive contributions of $\Od(1/\Gamma)$ to the conductivity. Let
us show this in more detail. First, we define a $k$-bubble
effective vertex $\vec{B}^{(k)}$ with $k\geq 1$ (see Figure
\ref{fig:bubbles}) exactly as we have done in (\ref{condladdern})
for the ladder vertices $\vec{\Lambda}^{(k)}$.  Given the
structure of the bubble vertices, one readily realizes that,
unlike the ladder ones,  $\vec{B}^{(k)}
(i\omega_n+i\nu_m,i\omega_n,\vec{p})$ is independent of the
 frequency and momentum of the external pion leg
 $(i\omega_n,\vec{p})$ if only constant four-pion vertices are
 considered. Therefore, the contribution of those diagrams to the
 conductivity vanishes by parity, for instance by making $\vec{p}_1\rightarrow
 -\vec{p}_1$ in the integral analogous  to (\ref{condladdern}). Thus,
 one needs derivative vertices to yield a nonvanishing
 contribution. In fact, following the parity argument, the only
 contributions of a $k+1$-bubble diagram which do not vanish by parity
 are those where all the four-pion vertices contain the factor
 $\vec{p}_{i}\cdot\vec{p}_{i+1}$ ($i=1,\dots,k$)
 where $p_i$ is the momentum running through the $i$-th bubble. We
 remark that we are considering only pion vertices  coming from
 ${\cal L}_2$, higher order vertices being suppressed by
 Weinberg's power counting and the same holds for bubbles made with more pion lines, say with
 vertices with more than four pions. Thus,  the bubble effective
 vertices satisfy a recurrence relation of the form:

\begin{widetext}
\ba \vec{B}^{(k)}(i\nu_m,\vec{p})=\ptint
\frac{\vec{p}\cdot\vec{p}_2}{f_\pi^2} \ \tsum\vec{B}^{(k-1)}
(i\nu_m,\vec{p}_2) G_\beta(i\omega_n,\modpt)
G_\beta(i\omega_n+i\nu_m,\modpt)\label{evbubble}\ea
\end{widetext}

In fact,  we should take into account that here can be also
neutral pion loops in bubble diagrams with more than two bubbles,
implying different vertex and combinatoric numerical factors in
front of $\vec{p}_{i}\cdot\vec{p}_{i+1}$ for the
$\pi^0\pi^0\pi^+\pi^-$, $\pi^+\pi^-\pi^+\pi^-$ and $(\pi^0)^4$
vertices. Thus, we really have to define two effective vertices
$\gamma\pi^+\pi^-$ and $\gamma\pi^0\pi^0$, the latter starting
with one bubble and one has then two coupled recurrence relations
between those vertices, instead of the single one in
(\ref{evbubble}). However, that merely complicates  the
calculation but does not change our main conclusion, namely that
bubble diagrams are subdominant. Hence, for simplicity, let us
show it for the case in which all $4\pi$ vertices  contribute
simply as $\vec{p}_{i}\cdot\vec{p}_{i+1}/f_\pi^2$ and all internal
lines are charged. Proceeding then as in section
\ref{sec:laddlowt}, we realize that the calculation is
particularly simple now since the vertex (\ref{evbubble}) is still
independent of the frequency of the external leg $i\omega_m$ and
therefore there is no pole contribution, unlike (\ref{evlt}).
Then, the Matsubara sum is just the same as that in (\ref{cond0}).
We get, after analytic continuation $i\nu_m\rightarrow \omega
+i\epsilon$ and to leading order in $1/\Gamma$ and
$\omega\rightarrow 0^+$:

\begin{widetext}
\ba \vec{B}^{(k)}(\omega,\vec{p})= \frac{i\omega}{T}\ptint
\frac{\vec{p}\cdot\vec{p_2}}{f_\pi^2}
\frac{n(E_2)\left[1+n(E_2)\right]}{4E_2^2
\Gamma_2}\vec{B}^{(k-1)}(\omega,\vec{p}_2)\equiv
\frac{i\omega\vec{p}}{f_\pi^2} F^{(k-1)}(\omega;T) \ea
\end{widetext}
for $k\geq 1$, where the last line holds by Euclidean covariance
and defines  the scalar function $F$, which satisfies then the
recurrence relation:

\be  F^{(k)}(\omega;T)=\frac{i \omega I(T)}{f_\pi^2}
F^{(k-1)}(\omega;T)=2 I(T)\left[\frac{i \omega
I(T)}{f_\pi^2}\right]^k \label{geom}\ee where $I(T)$ is the
  integral  appearing in the lowest order conductivity in
(\ref{sigma0}), namely $\sigma^{(0)}=4 e^2 I(T)$ and we have used
that $B_j^{(0)}=\Lambda_j^{(0)}=2p_j$ so that $F^{(0)}=2I(T)$.

From the above results, we note that the two main differences
between the bubble effective vertices and the ladder ones are,
first, that for small $\omega$, the contribution of the $k+1$-bubble
diagram to the conductivity is $\Od(\omega^k)$ and, second, that the
bubble effective vertices are not all real, $B^{(k)}$ is real  for
$k$ even and purely imaginary for $k$ odd, to leading order in
$1/\Gamma$. This confirms our discussion in section
\ref{sec:laddlowt} about the role of the different diagrams as
related to the ``scattering'' of the internal pion lines. The
one-bubble diagram amounts to the tree level contribution to the
scattering in ChPT which is always real. The first diagram giving a
nonvanishing contribution to the scattering spectral function is the
two-bubble one, but it vanishes as $\omega\rightarrow 0^+$ as
corresponds to the ${\cal U}$ channel.

The simple form of the bubble vertices allows to perform
explicitly their contribution to the conductivity, since they form
basically a geometric series according to (\ref{geom}). The fact
that bubble diagrams can be summed  is a well-established result
and holds also in other analysis of transport coefficients
\cite{jeon95}. In fact, we remark that in our case it is
consistent  to sum them, following  our power counting arguments,
since the bubble diagrams are neither reduced by chiral loops like
the ladder ones (all lines are double) nor by higher order pion
vertices if all pion vertices come from ${\cal L}_2$.

Thus, the contribution of the sum of the bubble diagrams to the
conductivity is of the form:

\begin{widetext}
\ba \sigma_{bub}&=&\frac{e^2}{3}\lim_{\omega\rightarrow 0^+}
\frac{1}{T}\pint
  \frac{n(E_p)\left[1+n(E_p)\right]}{4E_p^2 \Gamma_p}
\re\sum_{k=1}^\infty 2\vec{p}\cdot
\vec{B}^{(k)}(\omega,\vec{p})\nonumber\\
&=&2e^2\lim_{\omega\rightarrow 0^+}\re\sum_{k=1}^\infty
  F^{(k)}(\omega)
=\sigma^{(0)}(T)\lim_{\omega\rightarrow 0^+} \re \frac{i\omega
I(T)/f_\pi^2}{1-i\omega
I(T)/f_\pi^2}\nonumber\\&=&\Od(\omega^2)\sigma^{(0)}(T)
\label{bubsum}\ea
\end{widetext}
and the same conclusion, i.e,
$\sigma_{bub}=\Od(\omega^2)\sigma^{(0)}(T)$ is reached when
considering the full contributions of charged and neutral diagrams,
as commented before. We wish to stress that the leading order in the
bubble vertices comes from the ``pinching poles'' in products $G_R
G_A$ and carries the $\Od(\omega/\Gamma)$ contributions analyzed
above. Taking the next to leading order could give a nonvanishing
contribution as $\omega\rightarrow 0^+$ but with no $1/\Gamma$ terms
(the $I(T)$ integrals above) and therefore suppressed by the
ordinary chiral power counting of the loops so that, typically, the
$\omega I(T)$ factors in the sum (\ref{bubsum}) would be replaced by
an $\Od(p^2)$ contribution. Therefore, our conclusion is that the
bubble diagrams are subdominant for all temperatures within the
validity range of ChPT.

\section{Unitarity, high $T$ behaviour and photon production}
\label{sec:unit}

One of the conclusions of  the preceding sections is that the
behaviour of the conductivity with temperature is greatly influenced
by that of the thermal width, as expected from physical arguments.
At the same time, the thermal width in the dilute gas regime
(\ref{gammadg}) grows with $T$ according to the growing of the
partial waves and cross-section with energies. We have been using
the low-energy leading order in ChPT (\ref{sigmatotlo}) which grows
with energy $s$. However, we know that  partial waves should be
bounded in energy, as a consequence of unitarity. We recall that the
unitarity condition for the $S$-matrix translates into

\be \im t_{IJ}(s)=\sigma_{2p} (s)\vert t_{IJ}\vert^2
\label{unitpw}\ee for the elastic $\pi\pi$ scattering partial
waves ($s\geq 4m_\pi^2$ and below any inelastic threshold) where
$\sigma_{2p}(s)=\sqrt{1-4m_\pi^2/s}$ is the two-pion phase space.
The unitarity condition (\ref{unitpw}) implies that the modulus
and the phase $\delta_{IJ}$ (the phase shifts) of the partial
waves are related as $\vert t_{IJ}
\vert=\sin\delta_{IJ}/\sigma_{2p}$ which means in particular that
$\vert t_{IJ} \vert$ is bounded as energy increases, as commented
above.

The ChPT partial waves only satisfy unitarity perturbatively, for
instance $\im t_{IJ}^{(4)}=\sigma_{2p} \vert t_{IJ}^{(2)} \vert^2$
where the chiral expansion is
$t_{IJ}=t_{IJ}^{(2)}+t_{IJ}^{(4)}+\dots$. They are not bounded and
they cannot reproduce a resonant behaviour, like those appearing
in the $I=J=0$ (the $\sigma$ or $f_0(600)$) and $I=J=1$ channels
(the $\rho(770)$).  However, it is possible to construct
amplitudes satisfying exact unitarity and matching the low-energy
ChPT expansion. In doing so, we will follow the so called Inverse
Amplitude Method (IAM) \cite{iam} relying on its remarkable
success to describe  meson-meson scattering data up to 1 GeV and
to generate the low lying resonances \cite{gp02}. We remark that
through this unitarization procedures, physical resonances are
generated dynamically with masses and widths in agreement with
those listed in the Particle Data Book \cite{pdg}. Furthermore,
the unitarization procedure can be extended to finite temperature,
providing a consistent description of the thermal properties of
resonances, such as the widening of the $\rho$ in the thermal bath
\cite{dglp02}.  The IAM unitarized partial waves for $\pi\pi$
elastic scattering are simply given by
$t^{IAM}=[t^{(2)}]^2/(t^{(2)}-t^{(4)})$. The same formula is valid
at finite temperature, where  the fourth order thermal amplitudes
have been calculated in \cite{glp02}.

We are then interested in the effect of unitarity on our previous
results. The first important comment is that unitarity is not
expected to change our low $T$ results, where only the low-energy
region of the $\pi\pi$ scattering contributes. We expect more
significant corrections for $T$ near $m_\pi$. The effect of
unitarity on the pion width has already been discussed in
\cite{schenk93}. As it follows from our previous comments, the
unitarized width $\Gamma_p$ has a softer behaviour with $p$,
which, according to our arguments in section \ref{sec:hit} means a
larger contribution to the conductivity as $T$ increases, as
compared with the non-unitarized one.

We replace  then the unitarized IAM partial waves in
(\ref{sigmatot}) in order to obtain the unitarized width
(\ref{gammadg}) and conductivity (\ref{sigma0}). Several comments
are in order here. First, we are considering only the contribution
of the partial waves with lower angular momentum, i.e.,
$IJ=00,11,20$. Partial waves with $J>1$ are negligible for
$\sqrt{s}\leq$ 1 GeV \cite{gp02} and so they are for the
temperatures involved here. Second, we are not including the finite
temperature corrections to the partial waves calculated in
\cite{glp02}. The reason is that those are dilute gas corrections to
the partial waves and therefore would contribute at $\Od(n_B^2)$ to
the thermal width in (\ref{gammadg}), i.e, at the same order that
other contributions that we have neglected. However, we should bear
in mind that  thermal corrections to the amplitudes describe
 correctly important features such as the changes in the mass and
width of the thermal resonances, as it has been indeed observed in
Relativistic Heavy Ion Collisions, so that their effect could be
qualitatively important. Finally, we are using the same set of
chiral parameters for the $d=4$ lagrangian as in \cite{dglp02},
ensuring that the mass and width of the $\rho$ agree with the
$T=0$ values  in \cite{pdg}.

\begin{figure*}\includegraphics[scale=.65]{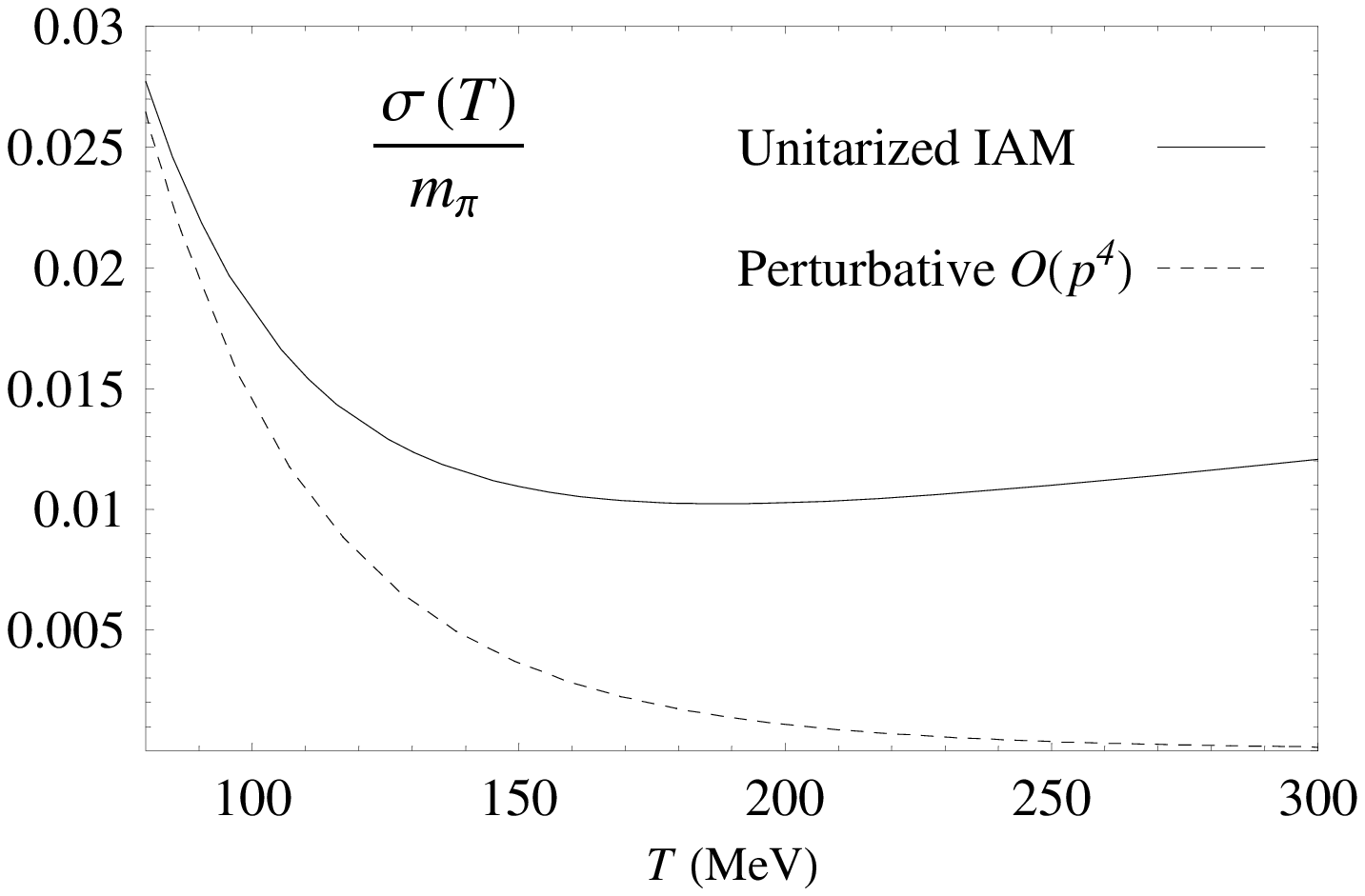}
 \caption{\rm
 \label{fig:unit} We show the result
 obtained in the dilute gas regime with the unitarized partial waves (solid line) compared
  to the perturbative ones up to $\Od(p^4)$ (dashed line).}
\end{figure*}

The results of the unitarized conductivity are showed in Figure
\ref{fig:unit}, where we compare with the non-unitarized one
including $\Od(p^4)$ terms in the partial waves, i.e.,
$t=t^{(2)}+t^{(4)}$. The most relevant feature is that for
temperatures about the pion mass, the curve for the conductivity
changes notably its behaviour with respect to the  perturbative one
for $T\gsim$ 100 MeV. As expected from our previous comments, the
unitarized conductivity decreases much more slowly with temperature.
In fact, extrapolating this result for high enough $T$, the
conductivity starts increasing with $T$, although very slowly as it
can be seen in Figure \ref{fig:unit}. Therefore, unitarity, and
ultimately the presence of resonances in the medium, is  responsible
for the change of behaviour from the low $T$ to the high $T$ phases.
In this sense, we notice that the main effect arising from
unitarization comes from the $00$ channel which has the vacuum
quantum numbers and is therefore more sensible to changes near the
chiral phase transition. We also remark that this softening of the
temperature behaviour when unitarization effects are taken into
account holds also for the shear viscosity \cite{dolla04}.

Although clearly we should not trust our approximations much beyond
$T\simeq$ 150 MeV,  it is interesting to note that all the high-$T$
analysis in the QGP phase shows a typical linear increasing
behaviour $\sigma\propto T$ \cite{amy00}, which follows simply from
high-$T$ dimensional analysis. Therefore, when we include unitarity
(which may not be the only relevant effect at these temperatures),
the conductivity slowly tends to its expected high-$T$ behaviour. In
fact, extrapolating our results to arbitrarily high temperature
yields a linear behaviour for $T\sim$ 600 MeV.

The high-$T$ QCD analysis are not applicable to temperatures close
to the transition, say $T\sim$ 200 MeV, where one should in
principle appeal either to lattice results or to phenomenological
models. The lattice calculation in \cite{gupta04} predicts
$\sigma\simeq 0.35 T$ for $T\simeq 1.5 T_c$. However, one should be
careful when interpreting results coming from the lattice in the
$\omega\rightarrow 0^+$ limit of the current-current correlators
\cite{resco}. In fact, there are notable discrepancies between
different lattice analysis. For instance, the conductivity
extrapolated from the dilepton rate  in \cite{karsch02} for $T\simeq
1.5T_c,3T_c$ vanishes identically, as pointed out in \cite{blage05}.
This is equivalent to state that the photon spectrum vanishes at
vanishing energy from the timelike region (see our comments in
section \ref{sec:photons}).

Our results show the behaviour from below the chiral phase
transition, although with our present level of approximation we
can only make qualitative statements at this point. On the one
hand, we predict a nonzero conductivity, which, as we discuss in
the next section, is compatible with having a sizable contribution
to the photon emissivity from the hadronic phase. On the other
hand, although unitarity makes the conductivity grow with $T$ as
we approach the chiral phase transition (note that the
nonunitarized result vanishes asymptotically) the values we obtain
are still much smaller than those in \cite{gupta04}. For instance,
we have $\sigma/T\simeq 0.007$ around $T=$ 200 MeV. Apart from
possible ambiguities in the lattice analysis, we should bear in
mind  that several effects that we have not accounted for in our
perturbative treatment, like resummation of ladder diagrams or the
contribution of heavier particles, become important at those
temperatures.

\subsection{Photon production}
\label{sec:photons}

Another possibility to extract physical information related to the
electrical conductivity is to consider the photon emissivity coming
from the hadronic phase. The photon differential rate emerging from
an equilibrated system is directly related  with the EM
current-current correlator as \cite{alam01}:

\be \omega\frac{d R_\gamma}{d^3 \vec{q}}=\frac{1}{8\pi^3}n_B(\omega)
\rho^{\mu}_{\mu}(\omega=\vert \vec{q} \vert) \label{rate}\ee

The Ward Identity $q^\mu\rho_{\mu\nu}=0$ implies
$\rho^0_0(\omega\neq 0,\vec{q}=0)=0$ so that the electrical
conductivity provides direct information about the vanishing energy
limit ($\omega\rightarrow 0^+$)  in the static region
\cite{gupta04,blage05} simply as $\omega dR_\gamma/d^3\vec{q}
(\omega\rightarrow 0^+,\vec{q}=0)=3T\sigma(T)/(4\pi^3)$, i.e, a
nonzero conductivity implies a constant value for the photon yield
near the origin\footnote{This result for the photon spectrum
extrapolates smoothly from the timelike static region
$(\omega,\vec{0})$ to the lightlike one. Remember that the  small
energy and momentum limits of the thermal correlators may give
different answers. We have checked explicitly  that to leading order
$\Delta^R_{00}(\omega\neq 0,\vec{0})=0$, according to the Ward
Identity, when a tadpole diagram contributing to
$\re(i\Delta^R_{00})$  is added to the diagram in Figure
\ref{fig:diagLO}. However, from
(\ref{loopcorrj})-(\ref{looptheralpha}) we find that
$\rho_{0}^{0}(\omega\rightarrow 0^+,\vert\vec{q}\vert\rightarrow
0^+)/\omega$ vanishes for slightly timelike photons but diverges for
spacelike ones, analogously to (\ref{condnogamma}).}.

 With this motivation in mind, we will compare  our results for the
conductivity for temperatures physically relevant in a
Relativistic Heavy Ion environment  with  some recent theoretical
and experimental analysis of the low-energy hadronic photon
spectrum. Among the theoretical works, one can find the virial
expansion approach followed in \cite{steele96}  with nonstrange
mesons only, extended in \cite{steele97} to include baryon
density. More recent analysis follow  the conventional Kinetic
Theory approach, where all possible photon-producing reactions of
the type $1+2\rightarrow 3+\gamma$ are evaluated,
 1,2,3 being meson degrees of freedom $\pi,\rho,K,...$
parametrized in an effective lagrangian with explicit resonances
included \cite{alam01,rawa99,turaga04}. The present experimental
data are described remarkably well for $\omega>$200 MeV when all the
relevant processes are evaluated and finite density effects are
accounted for \cite{turaga04}. However, it is worth mentioning that
recent analysis of the WA98 experiment \cite{wa98} at CERN have
produced data at lower transverse momentum, showing a slight
systematic enhancement with respect to theoretical predictions. At
those low energies, the hadronic gas contribution seems to dominate
over the QGP one.

 The low-energy rate with nonstrange mesons in
\cite{steele96}, which deals then with the same degrees of freedom
as our approach,  reaches a maximum value of about 3 $\times
10^{-5}$ GeV$^{-2}$fm$^{-5}$  (for $T=$150 MeV) at $\omega\sim$ 400
MeV and then drops to zero at the origin. When including baryons in
the same approach \cite{steele97} the behaviour is the same but the
maximum is at $\omega\sim$ 200 MeV and grows up to about 4 $\times
10^{-4}$ GeV$^{-2}$fm$^{-4}$. The results in \cite{rawa99} for the
hadronic gas  are similar to those in \cite{steele96,steele97}
except that the curve with baryon density decreases monotonically
for $\omega\geq$ 200 MeV. In \cite{turaga04}, more relevant channels
have been included, the most interesting result being that there are
also meson contributions that do not vanish when approaching the
origin, for instance those involving $\pi\rho a_1$ mesons. The
latter contributions  amount to almost twice more photons near the
origin that in the earlier work \cite{rawa99}. We remark that the
prediction in \cite{turaga04} is  the one used in the experimental
WA98 paper \cite{wa98} to compare with data and it remains below the
 lowest energy points. Using the unitarized width, we get for the vanishing energy photon
rate $3T\sigma/(4\pi^3)\simeq 3.7\times 10^{-3}$
GeV$^{-2}$fm$^{-4}$ at $T=$150 MeV. Therefore, we predict a
sizable value of the hadronic photon rate at the origin. A linear
extrapolation of the low energy curve ($\omega\simeq$ 200 MeV) in
\cite{turaga04} to the origin gives a value of about 2 $\times
10^{-3}$ GeV$^{-2}$fm$^{-4}$ at $T=$150 MeV. This is an indication
that our prediction lies in the correct range.

We will now try to establish a more direct comparison with
experimental data, bearing in mind  that our approach only gives
information about the value very close to the origin. One then has
to integrate the photon rate in (\ref{rate}) over space-time and
average over the photon rapidity \cite{turaga04} in order to
obtain the experimentally measured photon yield $\omega
dN_\gamma/d^3\vec{q}$ as a function of $q_T$, the component of the
photon momentum transverse to the collision axis in the laboratory
frame. Obviously, the results depend heavily on the hydrodynamical
space-time evolution for the conditions applicable to a particular
experiment. We will just give here rough estimates and for that
purpose we will neglect transverse flow and assume a simple
Bjorken's hydrodynamical description \cite{lebellac,petho02} so
that $\omega=q_T\cosh (y-\eta)$ with $y$ and $\eta$ the photon and
fluid rapidities respectively. Therefore, our value at
$\omega\rightarrow 0^+$ translates directly into the value of the
yield at $q_T\rightarrow 0^+$. As for the space-time integration,
we use the typical value \cite{lebellac,petho02} for the nuclear
transverse radius $R_A\simeq$ 1.3 fm $A^{1/3}\simeq$ 7.7 fm for
${}^{208} $Pb and change to proper time $\tau$ and rapidity $\eta$
coordinates. In the Bjorken's limit and for $q_T\rightarrow 0^+$,
there is no dependence with rapidity so that we have simply to
multiply by the expansion velocity
 \cite{petho02}
$\Delta \eta_{nucl}=2\mbox{arccosh} (\sqrt{s}/(2 A GeV)\simeq 10.1$
for the WA98 collision energy $\sqrt{s}=158$ A GeV.  With these
approximations,
 the photon yield at the origin is given by:

\be\omega \frac{dN_\gamma}{d^3\vec{q}} (q_T\rightarrow 0^+)\simeq\pi
R_A^2\Delta \eta_{nucl}\int_{\tau_i}^{\tau_f} \frac{3
T(\tau)\sigma(T(\tau))}{4\pi^3}\tau d\tau\label{yield}\ee

A crude estimate  is obtained  by assuming a purely thermal
hadronic phase, i.e., a constant temperature. Taking proper time
values $\tau_i=$ 1 fm/$c$ and $\tau_f=$ 13 fm/$c$ \cite{turaga04}
and $T=150$ MeV, this  gives $dN_\gamma/d^3\vec{q} (q_T=0)\simeq
5.8\times 10^2$. A more realistic approximation is  to take the
cooling law of an ideal gas $T(\tau)=T_i(\tau_i/\tau)^{1/3}$
\cite{lebellac} which is probably still rather crude for mesons at
moderate or high temperatures. Taking  $\tau_i=$ 3 fm/$c$, more
appropriate for the hadronic phase, $\tau_f=$ 13 fm/$c$ and $T_i=$
170 MeV, we get $T_f=$ 104 MeV which is of the order of the
freeze-out temperature. Inserting this law in (\ref{yield}) with
the unitarized conductivity in Fig.\ref{fig:unit} we obtain
$dN_\gamma/d^3\vec{q} (q_T=0)\simeq 5.6\times 10^2$ although this
result is very sensitive to variations in  temperature.

Taking the two points of smallest $p_T$ in \cite{wa98} and simply
extrapolating them to the origin with a straight line gives
$dN_\gamma/d^3\vec{q} (q_T=0)\simeq 5\times 10^2$. Therefore, our
results are  compatible with the recent data, in the sense of a
naive linear extrapolation from the origin and not forgetting  that
we are bordering the applicability range of our approach, in
addition to the many approximations performed to arrive to a number
comparable with experiment. In any case, our analysis would suggest
that purely thermal cuts in
 pion-pion annihilation (there is no $T=0$ real photon production from $\pi\pi$ annihilation
 for energies below $2m_\pi$) with a nonzero pion thermal width
  may be a relevant effect \footnote{We would get a vanishing contribution to the $\omega\rightarrow 0^+$
  photon spectrum with zero pion width and slightly timelike photons, from (\ref{condnogamma}).}.
  In this sense, it should be borne in mind that the pion thermal width
   is not considered in
\cite{steele96,steele97,alam01,rawa99,turaga04} since it does not
play any role at the energies of interest for the photon yield
considered in those works. On the other hand, our approach could
be too limited to describe correctly the effect  of some
resonances like the $a_1$. It is  even unclear whether our
prediction at small energies  has simply to be added to previous
hadronic ones, since we are extracting the photon rate from the
self-energy and not from individual processes. What is    more
clear in this respect is that the contribution analyzed here does
not come from baryonic sources.

\section{Conclusions}
\label{sec:conc}

In this work we have analyzed the DC electrical conductivity of a
pion gas at low temperatures within the framework of Chiral
Perturbation Theory. The nonperturbative nature of transport
coefficients is reflected in the need of including the thermal pion
width in the calculations, in order to avoid ``pinching poles''
singularities. Physically, this allows to account for the relevant
in-medium pion collisions and gives rise to the leading order
contribution in the inverse width consistently with Kinetic Theory.
The leading order in ChPT shows a decreasing behaviour with
temperature, behaving like $\sigma\sim\sqrt{m_\pi/T}$ for $T<<m_\pi$
 as expected from  the nonrelativistic limit of Kinetic Theory. For
higher temperatures, $\sigma\sim (m_\pi/T)^2$ for $T\lsim m_\pi$ if
unitarity corrections are not included in  elastic pion scattering.

A very important part of our analysis has been  the role of higher
order diagrams which, although naively suppressed, are enhanced by
powers of the inverse width. As it happens in other theories, the
dominant diagrams are uncrossed ladder ones, which in our case  can
be interpreted in terms of pion scattering of the internal lines. A
careful evaluation of the ladder diagrams shows that they can be
still considered perturbative in ChPT at low temperatures. This is
particularly important at very low temperatures $T<<m_\pi$ where the
nonperturbative contributions are larger. In that regime we have
been able to show exactly that ladder diagrams are perturbative, so
that they merely renormalize the coefficient of the $1/\sqrt{T}$
behaviour of the conductivity by subleading corrections in ChPT.
Although collinearly enhanced, our numerical analysis shows that
these corrections are typically around  5\%. As temperature
increases, ladder diagrams become more important, the most relevant
contributions coming from ladders with derivative vertices. At
temperatures near the chiral phase transition it is not clear that
ladder diagrams can be neglected, since ChPT is not applicable. In
fact, at those high temperatures we expect other effects to become
important, like the presence of kaon states.

Another important point concerns unitarity. We have showed that the
conductivity changes qualitatively its behaviour with temperature
for $T>$100 MeV as a consequence of implementing unitarity of the
partial waves  in the thermal width. This provides  a more physical
picture as far as the behaviour of partial waves with energy and the
presence of resonances in the thermal bath are concerned. The
unitarized conductivity increases slowly for increasing $T$, which
 seems to be consistent
 with lattice and
 analytical  analysis far beyond the transition point. In fact, the
 most important contribution in this respect comes from the  partial wave
 corresponding to the $f_0 (600)$ with  vacuum quantum numbers $I=J=0$.
 We have also considered some possible consequences of our results
 regarding the photon spectrum. The lack of precise experimental and
 lattice knowledge about the zero frequency limit, together with the
own limitations of our approach do not allow to draw very
quantitative conclusions. However, our analysis implies that there
should  be sizable effects for very low energy hadronic photon
production and this  result is consistent with  recent theoretical
low-energy analysis and compatible with naive extrapolations of
experimental data. In our opinion, this makes the analysis presented
here useful in order to understand some of the current discrepancies
regarding the low-energy results. However, this is just but the
first step and we will pursue a more detailed analysis
 in the near
future, including some effects discarded here that could be
important in more realistic situations, like the presence of kaon
states, nonzero pion and baryon chemical potentials or thermal
modifications of resonances.

Summarizing, despite its limitations as the temperature approaches
the phase transition, we believe that our present work provides a
useful and physically relevant example where the diagrammatic
approach to transport coefficients can be implemented in a
controlled way. More realistic  applications to physical observables
such as the photon spectrum are under way and will be reported
elsewhere.

 \begin{acknowledgments}We are grateful to R.F. \'Alvarez-Estrada, A. Dobado,
 F.J. LLanes-Estrada and J.M. Mart\'{\i}nez Resco for their
  useful comments. We also acknowledge financial support from the Spanish research projects
FPA2004-02602, BFM2002-01003, PR27/05-13955-BSCH, FPA2005-02327
and from the fellowship (D.F.F) BES-2005-6726 of the Spanish
F.P.I.  programme.
\end{acknowledgments}

\appendix*
\section{Thermal correlators}

\subsection{General Properties}

We review here some of the results needed throughout the text,
mostly to fix our conventions and notation. For a given,
rotationally invariant,  correlator $\Delta (z,\modq)$ analytical
for any complex $z$ off the real axis, it is convenient to
introduce its spectral function as:

\be \Delta (z,\modq)=\int_{-\infty}^{\infty} \frac{d\omega'}{2\pi}
\frac{\rho (\omega',\modq)}{\omega'-z} \ee

The advanced and retarded correlators are then given by \footnote{We
note that our convention for the retarded and advanced correlators
coincides with that in \cite{lebellac} but differs from
\cite{valle02} by the multiplying $\mp i$ factors.}:

\be \Delta^{R,A}(\omega,\modq)=\mp i \Delta (z=\omega\pm
i\epsilon,\modq)
\label{retadvdef} \ee
 where
$\omega\in\IR$. Note that, with our definitions
$\rho=2\re\Delta^R=2\im(i\Delta^R)$.

 On the other hand, the
imaginary-time correlator is given by the
 values on the discrete Matsubara frequencies:

 \be
\Delta_\beta (i\omega_n,\modq)=\Delta (z=i\omega_n,\modq)
\label{itdef}
 \ee

 When performing a perturbative calculation in the imaginary-time
 formalism, one can express the final result as an analytic function
 off the real axis, once all the sums over internal frequencies have
 been performed. Then, the imaginary-time result is continued
 analytically to external real frequencies simply from
 (\ref{retadvdef}) and (\ref{itdef}) as:

 \be
\Delta^{R,A} (\omega,\modq)=\mp i\Delta_\beta (i\omega_n\rightarrow
\omega\pm i\epsilon,\modq)
 \ee

The usual case is that when the correlator is a two-point function,
like the field $\phi$ itself or a conserved current. In that case
one has in position space:

\ba G^{R,A} (t,\vx)&=&\pm\theta(\pm t) <[\phi(t,\vx),\phi(0)]>\nonumber\\
G_\beta(\tau,\vx)&=&<T_C\phi(-i\tau,\vx)\phi(0)>\nonumber\\
\rho(t,\vx)&=& <[\phi(t,\vx),\phi(0)]>\ea where the brackets denote
thermal expectation values, $t\in\IR$ and $T_C$ denotes time
ordering along the imaginary-time contour $C$ defined as the
straight line  $\tau\in[0,\beta]$. Note that, from the above
equations, $\rho(\omega,\modq)=-\rho(-\omega,\modq)$, as expected
for  scalar spectral functions.

\subsection{Loop integrals}
In the analysis of the conductivity, we find loops involving the
imaginary-time correlators:
\begin{widetext}
\be I(i\omega_m,\modQ;k)=i\tsum\pint
\frac{\modp^{2k}}{\left[\omega_n^2+\modp^2+m^2\right]
\left[(\omega_n+\omega_m)^2+\modpcapq^2+m^2\right]}\label{loopcorr}\ee
\end{widetext}
with $k=0,1$.

Performing the Matsubara sum in the usual way \cite{lebellac}, one
arrives to
 a function analytical in $i\omega_m$ off the real axis, so that
 we find for  the retarded correlator:
\begin{widetext}
\ba \im I^R(\omega,\modQ;k)= \pi\sum_{s_1,s_2=\pm
1}\pint\modp^{2k}\frac{s_1 s_2}{4 E_p E_{p+Q}} \left[1+n \left(s_1
E_p\right) + n\left(s_2
E_{p+Q}\right)\right]\nonumber\\\times\delta\left[\omega-s_1
E_p-s_2 E_{p+Q}\right] \ea
\end{widetext}
where $E_q^2=\vert\vec{q} \vert^2+m^2$ and $n_B(x)=\left(e^{\beta
x}-1\right)^{-1}$ is the Bose-Einstein distribution function. The
two $s_1=s_2$ contributions require $s\geq 4m^2$
($s=\omega^2-\modQ^2$) and give the unitarity cut appearing in the
analysis of thermal scattering \cite{glp02}. Performing the
angular integrations with the delta function, it can be written
as:
\begin{widetext}
\be \left[\im
I^R(\omega,\modQ;k)\right]_{unit}=\frac{\sgn(\omega)\theta(s-4m^2)}{16\pi\modQ}
\int_{y_-}^{y_+}dy
\left(y^2-m^2\right)^k\left[1+n_B(y)+n_B(\vert\omega\vert-y)\right]
\ee
\end{widetext}
where $y_{\pm}=(\vert\omega\vert\pm\modQ\sigma_{2p}(s))/2\geq m$ for
$s\geq 4m^2$ and $\sigma_{2p}(s)=\sqrt{1-4m^2/s}$ is the phase space
of two particles of equal mass $m$.

The case $k=0$ can be solved in terms of elementary functions:
\begin{widetext}
\ba \left[\im
I^R(\omega,\modQ;0)\right]_{unit}=\frac{\sgn(\omega)\theta(s-4m^2)}{16\pi}
\left[\sigma_{2p}(s)+\frac{2
T}{\modQ}\log\left[\frac{1-\exp(-y_+/T)}{1-\exp(-y_-/T)}\right]\right]
\label{unitcut0}\ea
\end{widetext}

 On the other hand, in the limit
$\modQ\rightarrow 0^+$ (center of mass limit in the scattering case)
we find:

\begin{widetext}
\ba \left[\im I^R(\omega,\modQ\rightarrow
0^+;k)\right]_{unit}=\frac{\sgn(\omega)\theta(s-4m^2)}{16\pi}\sigma_T
(\omega)\left[\left(\frac{\vert\omega\vert}{2}\right)^2-m^2\right]^k
\ea
\end{widetext}
where $\sigma_T
(\omega)=\sigma_{2p}(\omega^2)\left[1+2n_B(\vert\omega\vert/2)\right]$
is the two-particle thermal phase space, which can be interpreted in
terms of enhancement and absorption in the thermal bath
\cite{glp02}.

Let us analyze now the two $s_1=-s_2$ contributions. First, it is
not difficult to see that these two integrals are identical and
contribute only for $s\leq 0$. Unlike the unitarity cut, this cut is
purely thermal, i.e, is only present for $T\neq 0$. Integrating the
delta function, their contribution to $\im I^R$ is:

\begin{widetext}
\ba \left[\im
I^R(\omega,\modQ;k)\right]_{ther}&=&-\frac{\sgn(\omega)\theta(-s)}{16\pi\modQ}\int_{y_0}^\infty
dy
\left[n_B(y+\vert\omega\vert)-n_B(y)\right]\nonumber\\
&\times&\left[(y^2-m^2)^k+ ((y+\vert\omega\vert)^2-m^2)^k\right] \ea
\end{widetext}
where $y_{0}=-y_-=(-\vert\omega\vert+\modQ\sigma_{2p}(s))/2\geq m$
for $s\neq 0$. For $k=0$:

\begin{widetext}
\ba \left[\im
I^R(\omega,\modQ;0)\right]_{ther}=\frac{T}{8\pi\modQ}\sgn(\omega)\theta(-s)
\log\left[\frac{1-e^{-(y_0+\vert
\omega\vert)/T}}{1-e^{-y_0/T}}\right]\label{thercut0}\ea
\end{widetext}

Now, taking the limit $\modQ\rightarrow 0^+$ with $m$ and $T$ fixed
implies necessarily $\omega\rightarrow 0^+$ since this is a
space-like contribution. More precisely,
$y_0\simeq\frac{m\modQ}{\sqrt{-s}}$, so that:
\begin{widetext} \be \left[\im I^R(\omega\rightarrow 0^+,\modQ\rightarrow
0^+;k)\right]_{ther}=-\frac{1}{8\pi}\theta(-s) \frac{\omega}{\modQ}
\left[F_k\left[\omega^2/\modQ^2;T\right]+
\Od\left(\frac{\vert\omega\vert}{m},\frac{\vert\omega\vert}{T}\right)\right]
\label{cortetersoft} \ee
\end{widetext}
 where: \be
F_k(x;T)=\int_{\frac{m}{\sqrt{1-x}}}^\infty dy (y^2-m^2)^k n_B'(y)
\label{Fdef}\ee for $0\leq x< 1$. Note that $F_k(x)<0$,
$\lim_{x\rightarrow 1^-} F_k(x)=0$ and that the small energy and
momentum limits (\ref{cortetersoft})  do not coincide necessarily
with taking directly $\modQ=0$ with $\omega\neq 0$, which gives a
vanishing contribution due to the step function.

Following similar steps as above, it is not difficult to obtain
the imaginary part of loop integrals with more powers of
$i\omega_n$ in the numerator:

\begin{widetext}
\be J_\alpha(i\omega_m,\modQ)=i\tsum\pint
\frac{(i\omega_n)^\alpha}{\left[\omega_n^2+\modp^2+m^2\right]
\left[(\omega_n+\omega_m)^2+\modpcapq^2+m^2\right]}\label{loopcorrj}\ee
\end{widetext}
with  $\alpha$ a positive integer or zero. We have analyzed the
$\alpha=0$ case previously. For arbitrary $\alpha$ we get:

\begin{widetext} \be \left[\im
J^R_\alpha(\omega,\modQ)\right]_{unit}=\frac{\left[\sgn(\omega)\right]^{\alpha+1}\theta(s-4m^2)}{16\pi\modQ}
\int_{y_-}^{y_+}dy
(-y)^\alpha\left[1+n_B(y)+n_B(\vert\omega\vert-y)\right] \ee \ba
\left[\im
J^R_\alpha(\omega,\modQ)\right]_{ther}=-\frac{\left[\sgn(\omega)\right]^{\alpha+1}
\theta(-s)}{16\pi\modQ}\int_{y_0}^\infty dy
\left[y^\alpha+(-1)^\alpha(y+\vert\omega\vert)^\alpha\right]\nonumber\\\times
\left[n_B(y+\vert\omega\vert)-n_B(y)\right]\label{looptheralpha}\ea
\end{widetext}


\end{document}